\documentclass[pra,preprint]{revtex4}
\usepackage{epsfig,graphicx}
\usepackage{amsfonts}
\usepackage{bm}% bold math

\begin{document}
\title{Electronic excitations in atomic clusters: beyond dipole plasmon}
\author{V.O. Nesterenko$^1$, P.-G. Reinhard$^2$, and W. Kleinig$^{1,3}$}
\date{\today}
\address{$^{1}$
Bogoliubov Laboratory of Theoretical Physics, Joint Institute for Nuclear
Research, Dubna, Moscow region, 141980, Russia}
\address{$^{2}$
Institut f\"ur Theoretische Physik II, Universit\"at Erlangen, D-91058,
Erlangen, Germany}
\address{$^{2}$
Technische Univirsitat Dresden, Inst. f\"ur Analysis,
   D-01062, Dresden, Germany}

\begin{abstract}
  Multipole electron modes beyond the Mie plasmon in atomic
    clusters are investigated
    %from a theoretical perspective
    within the
    time-dependent local density approximation theory (TD-LDA). We
    consider the origin of the modes, their connection with basic
    cluster properties and possible routes of experimental
    observation. Particular attention is paid to infrared magnetic
    orbital modes, scissors and twist, and electric quadrupole mode.
    The scissors and twist modes determine orbital magnetism of
    clusters while the electric quadrupole mode provides direct access
    to the single electron spectra of the cluster. We examine
    two-photon processes (Raman scattering, stimulated emission
    pumping and stimulated adiabatic Raman passage) as the most
    promising tools for experimental investigation of the modes.
\end{abstract}

%\pacs{PACS numbers: 36.40.Cg; 42.62.Fi; 42.50.Hz}
\maketitle

\arraycolsep1.5pt

\narrowtext

\section{Introduction}
Besides the dominant dipole Mie plasmon, many other electron modes,
both electric and magnetic, can exist in atomic clusters \cite{Ne_SN}.
These modes are known in diverse many-body systems (atomic nuclei,
quantum dots, dilute gas of trapped fermionic and bosonic atoms, etc).
It is a demanding problem to observe them in atomic clusters. The
multipole electron modes represent an essential part of electron
dynamics in both collective and electron-hole domains.
They are connected with basic cluster properties and thus can serve as
an effective tool for their investigation.

We will pay the main attention to three infrared modes: scissors magnetic
dipole (M1), twist magnetic quadrupole (M2) and electric quadrupole (E2). These
modes are connected basic properties of atomic clusters. Besides, they are the
strongest spin-saturated modes beyond the Mie plasmon.

The scissors and twist modes are of orbital magnetic character
\cite{LS_M1,LS_ZPD,prl_M1,twist_prl_M2}. In general, the scissors mode exists only in
deformed systems. It dominates the Van Vleck paramagnetism \cite{LS_ZPD,M1_ne_EPJD_03}
and can result in dia-para magnetic anisotropy
in particular light atomic clusters \cite{M1_ne_EPJD_03}. The twist mode
is the strongest magnetic orbital mode in spherical clusters where the
scissors mode vanishes \cite{twist_prl_M2}. It is mainly generated
by  transitions between electron levels with maximal orbital moments. Altogether,
the scissors and twist modes are fundamental sources of orbital magnetism in
spin-saturated clusters. In deformed clusters, the scissors M1 is coupled to
electric quadrupole E2 mode.

The infrared E2 mode is most interesting in two particular cases of
deformed clusters: free light deformed clusters and embedded oriented
rods (strongly prolate large clusters).  In the first case, the E2
mode is reduced to a few electron-hole excitations driven by cluster
deformation. As was recently demonstrated, these excitations allow to
determine the mean field spectra of light clusters
\cite{stirap_Ne_PRA}.  Being sensitive to cluster structure, these
spectra can deliver important information on diverse cluster features.
And last, but not least, the infrared electron-hole quadrupole
excitations have a good chance to be observed in particular two-photon
processes.

Infrared E2 modes in oriented silver rods embedded into glass matrices
\cite{rods_93,rods_98} represent another useful example of non-dipole
electron motion in clusters. These modes are collective and in
principle can be studied in Raman scattering (experiments of this kind
are in progress \cite{Duval}). Because of the extreme axis ratio, rods
represent a unique sample for investigation of both E2 and M1
(scissors) collective motion.

Being non-dipole, the infrared multipole states cannot be populated by
one-photon transitions which excite exclusively dipole modes. Thus one
has to use two-photon processes (TPP) where the target state is
populated via an intermediate dipole state by two (absorption and
emission) dipole transitions. The dipole plasmon or isolated dipole
states can serve as the intermediate state.  Some of the processes,
Raman scattering (RS), stimulated emission pumping (SEP), and
stimulated adiabatic Raman passage (STIRAP), are widely used in atomic
and molecular spectroscopy to populate non-dipole modes but their
applications to clusters are still very limited and, at a first
glance, even questionable. The problems are caused by particular
clusters properties (dense spectra, broad level structures,
harmonicity of collective modes, short lifetimes, non-radiative decay
channels, etc) which can hamper TPP. Nevertheless, we will show that
TPP listed above can be applied to clusters and deliver valuable
information on cluster properties (see preliminary discussion and
first estimations in \cite{stirap_Ne_PRA}).  Specific requirements to
TPP applications will be considered in detail.

The manuscript provides a survey of explorations from a theoretical
perspective. To that end, we employ the microscopic time-dependent
local-density-approximation (TDLDA) theories in the linear regime
\cite{Ne_AP} and beyond it \cite{Cal,Reibook}.  The main aim of our
study is to outline non-dipole electron modes in clusters and
encourage their investigation in TPP experiments.

The paper is outlined as follows.  In Sec. \ref{sec:review}, we give a
brief overview of infrared multipole modes, clarify their origin and
connection with basic cluster features.  In Sec. \ref{sec:exp}, the
two-photon processes RS, SEP and STIRAP are inspected as possible
routes of experimental study of the modes. Mainly free light deformed
clusters are considered. A summary is given in Sec.~\ref{sec:summary}.

\section{Hierarchy of multipole modes}
\label{sec:review}

\subsection{Classification scheme}

Multipole modes are oscillations of multipolarity $\lambda\mu$ which
are typical for finite many-body systems. It is useful to distinguish
collective and non-collective multipole modes. In the first case, one
deals with a superposition of many elementary excitations which form a
coherent motion while in the second case the mode represents a mixture
of only a few elementary excitations. In mesoscopic systems with
$10-10^4$ particles, we have an intermediate (and maybe the most
complicated) situation when a large number of small elementary
components contribute coherently and thus form the collective part of
the mode while a few dominant components determine its particle-hole
structure. Just such systems will be considered in the present study.
More specifically, we will deal with metal clusters where valence
electrons move freely in a common mean field. The single-particle
levels of the mean field are bunched into quantum shells
\cite{deH93,Bra93} which constitute a key point in understanding the
nature of multipole modes and their classification.

A useful sorting scheme for single-particle levels of valence
electrons in metal clusters is provided by the three-dimensional
harmonic oscillator \cite{Cle,deH93,Bra93}. The levels of a spherical
cluster are sorted in perfectly degenerate bunches (major quantum
shells) which are characterized by the principle quantum number ${\cal
  N}=0,1,2,...$.  The shells are separated by appreciable energy gaps
and every shell involves only states of the same space parity
$\pi=(-1)^{\cal N}$.  This oscillator picture is well fulfilled in
light clusters and provides still a good approximation in medium and
heavy ones. In axially deformed clusters, the single-electron levels
are characterized by Nilsson-Clemenger quantum numbers $\nu=[{\cal
  N}n_z \Lambda ]$ where the principle shell number ${\cal
  N}=n_z+2n_r+\Lambda$ is expressed through the numbers of nodes in
radial ($n_r$) and symmetry axis ($n_z$) directions and projection
$\Lambda$ of the orbital moment onto the symmetry axis \cite{Cle}.

Following this scheme, excitations of valence electrons are
characterized by the $\Delta {\cal N}$ value, the difference in shell
for the dominant $1eh$ jumps. Quadrupole excitations have even parity
and are collected into the branches $\Delta{\cal N} =0$ and 2. The
excitations with larger $\Delta{\cal N}$ are weak and can be
neglected.  The branch $\Delta{\cal N} =0$ has low excitation energy.
It exists only in deformed systems (with partly occupied valence
shell) and vanishes in spherical systems (with fully occupied valence
shell).  In the next subsections, we will use this sorting scheme for
explanation of the origin and properties of quadrupole E2, scissors M1
and twist M2 modes.

%%%%%%%%%%%
% Figure 1
%%%%%%%%%%%%
\begin{figure}
\includegraphics[height=10cm,width=7cm]{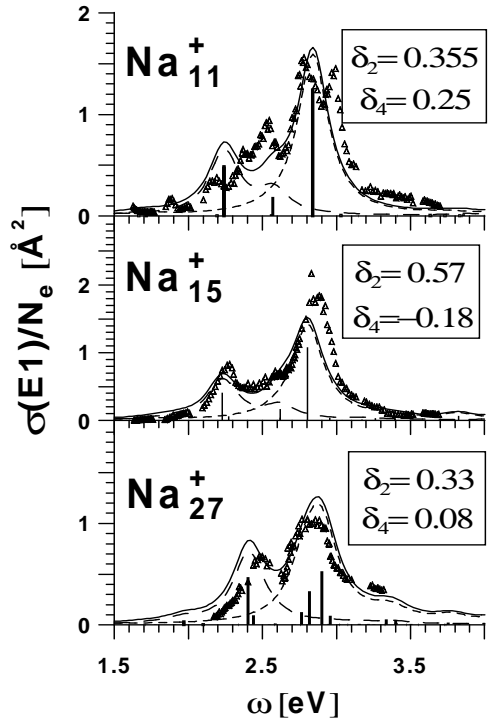}
%\centerline{\epsfig{figure=fig11.ps,width=9cm,clip=}}
\caption{\label{fig:e1_plasmon}
Photoabsorption dipole cross section in light axially deformed Na clusters.
Quadrupole and hexadecapole deformation parameters are indicated in boxes.
RPA results are given as vertical bars (in eV $\AA^2$) and as a strength
function smoothed by the Lorentz weight with the averaging parameter 0.25 eV.
Separate contributions to the strength function from the  $\lambda\mu =$10 and 11
dipole branches (the latter is twice stronger) are shown by dash curves.
The experimental data  (triangles) from \protect\cite{SH} are given for the comparison.
}
\end{figure}

\subsection{Calculation scheme}

The electron cloud is described by density-functional theory at the
level of the local-density approximation (LDA) for the ground state
and time-dependent LDA (TDLDA) for the excitations, using actually the
functional of \cite{GL}.  The ionic background of the cluster is
approximated by the soft jellium model allowing for quadrupole and
hexadecapole deformations \cite{Mon95b,Reibook}. In the analysis given
in this section, the infrared modes stay in the regime of small
amplitudes. We thus employ the linearized TDLDA, often called the
random-phase-approximation (RPA). The actual implementation for
axially symmetric clusters is explained in \cite{Ne_AP}. The
reliability of the method has been checked in diverse studies of the
Mie plasmon in spherical \cite{Ne_EPJD_98} and deformed
\cite{Ne_AP,Ne_EPJD_02} clusters. As an example, Fig.
\ref{fig:e1_plasmon} demonstrates quite good agreement of our results
with the experimental data \cite{SH} for the dipole plasmon in light deformed
clusters at room temperature.

The proper choice of the light clusters to be studied is very important.
i) The clusters should be small enough to possess a dilute and non-collective
infrared spectrum. Only then the spectrum can be resolved and unambiguously
related to the single-particle levels.
ii) Since infrared modes are mainly induced by cluster deformation
(see discussion in the next subsection), the clusters with a strong
deformation (either prolate or oblate) are desirable. The simplest
case of axial shape is most suitable for the analysis.
iii) Shape isomers exhibit different single-electron spectra
\cite{Ne_AP,Ne_EPJD_02,Kuemmel}, which can smear out the
low-energy spectral lines. The heavier clusters, the more
isomers \cite{Ne_AP,Ne_EPJD_02}. So, light
clusters with one dominant equilibrium shape are preferable.
Between them, we should choose the clusters whose ground state
and first isomers have the similar (prolate or oblate) shape.
Thus we will minimize the spectral blurring.
iv) The jellium approximation is certainly rough for description of
details in light clusters, which establishes a lower limit for the
cluster size. The results for the smallest samples may not reach a
quantitative level, but they are still useful for the first
consideration. As is demonstrated in Fig.  \ref{fig:e1_plasmon}, even
in the lightest clusters the jellium TDLDA sufficiently well
reproduces the basic characteristics of the Mie plasmon (average
energy, principle gross-structure, magnitude of the deformation
splitting of the resonance), see for discussion
\cite{Ne_EPJD_02,stirap_Ne_PRA}. Such accuracy suffices for our
present survey. For this reason, as well as for the sake of
simplicity, our study will exploit the jellium approximation.

\subsection{Quadrupole mode}

\begin{figure*} % Fig. 2
%\centerline{\epsfig{figure=fig1.eps,width=13cm,clip=}}
\includegraphics[width=13cm]{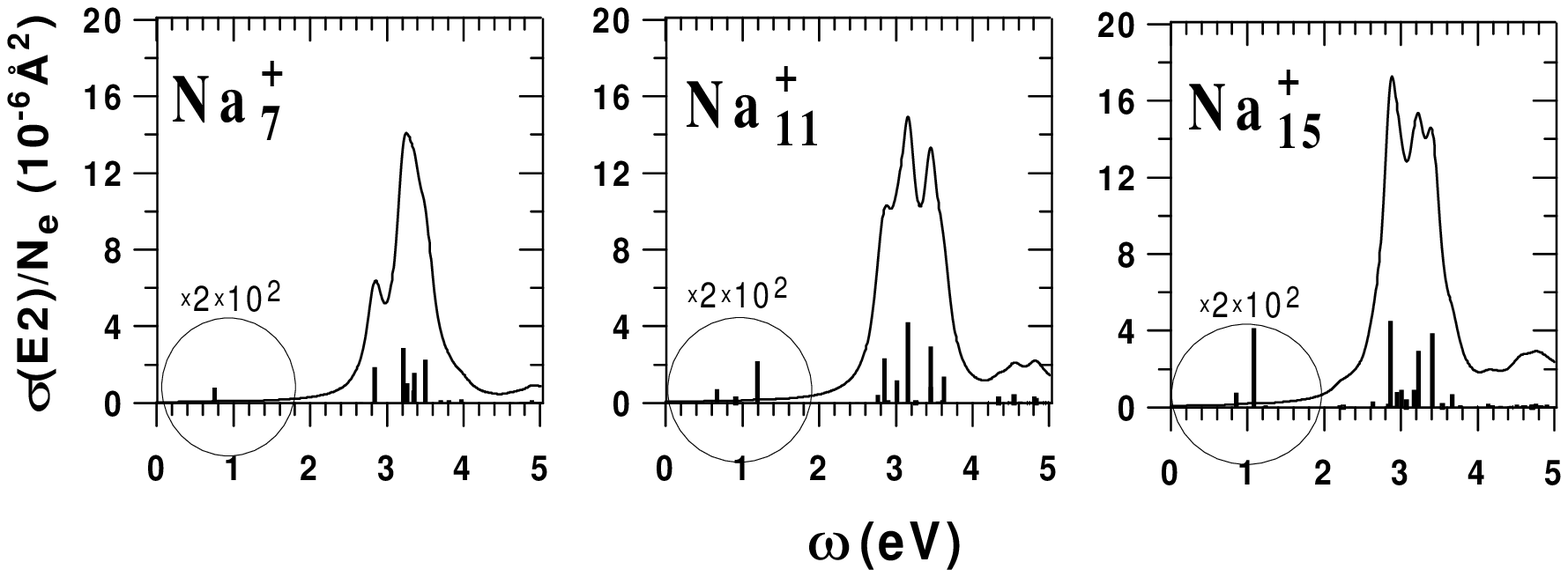}
\caption{\label{fig:fig2}
Quadrupole photoabsorption  in light deformed clusters
Na$_7^+$, Na$^+_{11}$ and Na$^+_{15}$. The results are given as bars for every
discrete state and as smooth strength functions obtained by folding with a
Lorentzian of width $\Delta =$0.25 eV. The weak infrared part of the strength
(enclosed by the circles at 0.5-1.5 eV) is rescaled by the factor $2 \cdot
10^2$. }
\end{figure*}

We start our analysis with quadrupole excitations in light deformed free sodium
clusters. The hierarchy of E2 modes is illustrated in Fig. \ref{fig:fig2} in
terms of normalized quadrupole photoabsorption \cite{Ne_AP}
$\sigma (\rm E2, \omega)/N_e \sim \sum_j\langle j|r^2Y_{2\mu}|0\rangle^2 \omega_j^3
\eta(\omega_j-\omega)$
calculated within RPA (here $\omega_j$ and is the energy of RPA $j$-state
and $\eta$ is the Lorentz weight function). Though E2 modes are
not observed in the photoabsorption, it is instructive for a first overview.
Let us first consider the strong quadrupole resonance appearing at high
frequencies in the range 2-4 eV.  In heavy clusters this resonance is
associated with the quadrupole plasmon. The resonance is mainly formed by E2
transitions over two major shells ($\Delta {\cal N} =2$).  It exists in
clusters of any shape and exhausts most of the quadrupole strength. Though the
resonance is energetically close to the dipole Mie plasmon, it can be
discriminated by means of angular-resolved electron energy-loss spectroscopy
(AR-EELS) at electron scattering angles $\sim 6^{\circ}$ \cite{Ger}.

%%%%%%%%%%%%
% Figure 2 %
%%%%%%%%%%%%
%%%%%%%%%%%%
% Figure 3 %
%%%%%%%%%%%%
\begin{figure} % Fig. 3
\centerline{\epsfig{figure=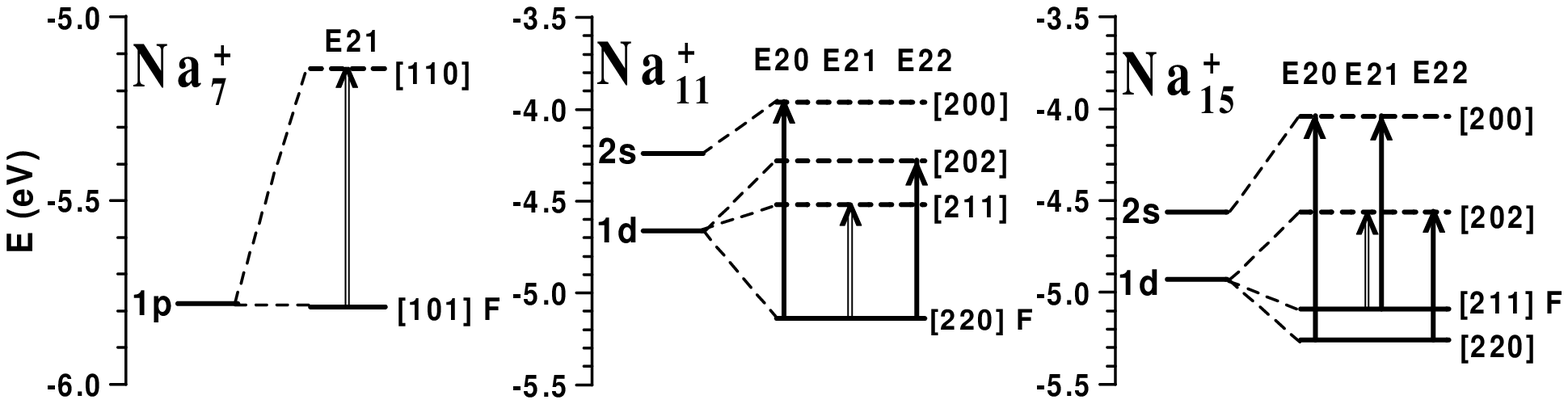,width=13cm,clip=}}
\caption{\label{fig:fig3} The electron level schemes for Na$_7^+$, Na$^+_{11}$
and Na$^+_{15}$ in the spherical limit (left) and at the equilibrium
deformation (right). Occupied and unoccupied single-particle levels are drawn
by solid and dash lines, respectively. The Fermi (HOMO) level is marked by
index F. Arrows depict the possible low-energy hole-electron $E2\mu$
transitions. Double arrows represent transitions of the scissor nature between
the members the deformation multiplet. }
\end{figure}

In the present study, we are not so much interested in the quadrupole
plasmon but in non-collective infrared quadrupole modes which can
deliver important information about the electron single-particle
spectrum near the Fermi (=HOMO) level. The infrared spectra are
associated with the low-energy $\Delta {\cal N} =0$ branch created by
E2 transitions inside the valence shell. Being of the $\Delta {\cal N}
=0$ origin, most of these spectra can exist only in clusters with {\it
  partly} occupied valence shell, i.e. in deformed clusters. The
deformation splits the infrared quadrupole mode into branches
$\lambda\mu=$20, 21 and 22.
In Fig. 1, the infrared modes reside at 0.5-1.5 eV.  As compared with the quadrupole
plasmon, the infrared spectrum is very dilute. It is represented only by
a few well separated levels. This prevents mixing of $1eh$ configurations by the
residual interaction and creation of collective states.  The infrared quadrupole
modes persist to keep their $1eh$ nature.  As is seen from
Fig. 1, they have very weak quadrupole strength in the photoabsorption.
But they may be accessible in two-photon processes.

%The modes can be easily identified as particular $1eh$ configurations.
Fig. \ref{fig:fig3} shows single-particle levels and
$E2\mu$-transitions inside the valence shells in Na$^+_7$,
Na$^+_{11}$, and Na$^+_{15}$. The levels can be characterized by
Nilsson-Clemenger quantum numbers \cite{Cle} $\nu=[{\cal N}n_z\Lambda
]$. The $1eh$ pairs corresponding to the transitions read $\{
[101]-[110]\}_{21}$ in Na$^+_7$, $\{ [220]-[200]\}_{20}$, $\{
[220]-[211]\}_{21}$, $\{ [220]-[202]\}_{22}$ in Na$^+_{11}$, and $\{
[220]-[200]\}_{20}$, $\{ [211]-[202]\}_{21}$, $\{ [211]-[200]\}_{21}$,
$\{ [220]-[202]\}_{22}$ in Na$^+_{15}$.  Following this scheme, the
infrared modes in Fig. \ref{fig:fig2} can be unambiguously identified
as particular $1eh$ configurations. The modes have different
photoabsorption strengths depending on value of their E2-transition
matrix element $\langle 1eh|r^2Y_{2\mu}|0\rangle$.
The RPA calculations which allow in principle any composition of of
states confirm that the infrared modes are indeed almost pure $1eh$
states. In every mode the dominant $1eh$ component typically attains
$99-100\%$.

The infrared modes provide valuable information about cluster's
properties.  As is seen from \ref{fig:fig3}, most of the transitions
connect the levels arising due to deformation splitting. The
corresponding infrared modes are determined by the deformation and
vanish at the spherical shape. Thus they deliver information on the
deformation splitting of the electron levels in the HOMO-LUMO region.
Besides, being combined with photoemission data for the spectra of
occupied electron states (see e.g. \cite{issen}), the infrared $1eh$
modes immediately yield energies of unoccupied electron states. As a
result, one can get the complete electron spectrum in the HOMO-LUMO
region. This spectrum is sensitive to different cluster features
(interplay of equilibrium and isomer cluster's shapes, ionic
structure, correlations, temperature effects, etc.) and so can serve
as an effective tool for investigation of these features.

Finally, it is worth noting that in deformed systems the electric and
magnetic modes with the same projection $\mu$ and space parity ${\pi}$
are mixed. The mixture of electric E21 and magnetic orbital M11
excitations is especially interesting as it provides access to the
orbital M1 scissors mode \cite{LS_M1,prl_M1,M1_ne_EPJD_03}.  The
properties of this mode are sketched in the next subsection.

\subsection{Scissors mode}

The scissors mode (SM) is a general collective flow already predicted
or found in different finite quantum systems (atomic nuclei
\cite{Iu_M1,richter}, quantum dots \cite{QD_M1} and ultra-cold
superfluid gas of fermionic atoms \cite{FA_M1}). Besides, this mode is
used as an indicator of Bose-Einstein condensate in the dilute gas of
trapped Bose atoms \cite{BE_M1,BEexp_M1}. In atomic clusters, the SM
was predicted \cite{LS_M1,prl_M1} but, in spite of some attempts
\cite{Duval_PRB_01}, not yet observed. Obviously, the SM
demonstrates a universal character.  It is pertinent to any finite
systems with two features in common: broken spherical symmetry
(deformation) and a two-component nature. In the systems listed above,
these components are neutrons and protons in nuclei, valence electrons
and ions in atomic clusters, electrons and surrounding media in
quantum dots, atoms and the trap in dilute Fermi and Bose gases. The
SM features were recently reviewed in \cite{M1_ne_EPJD_03}.

\subsubsection{Macroscopic view}

\begin{figure}
%\centerline{\epsfig{figure=view1.eps,width=8cm}}
\centerline{\epsfig{figure=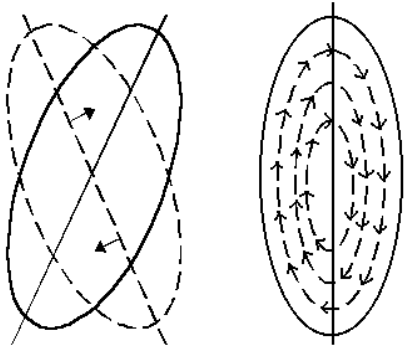,width=6cm,clip=}}
\caption{\label{fig:view}\sl
Macroscopic view of scissors mode : rigid rotation \cite{Iu_M1} (left),
and rotation within a rigid surface \cite{LS_ZPD} (right).}
\end{figure}

The macroscopic collective nature of SM can be illustrated by the
geometrical model first proposed for atomic nuclei \cite{Iu_M1}. For
clusters, this model exhibits SM as scissors-like small-amplitude
oscillations of valence electrons versus the ions, both assumed to
form distinct spheroids (see left part of Fig. \ref{fig:view}). Hence
the name {\it scissors} mode. Obviously, such a mode can exist only in
deformed systems.

Following the alternative view of \cite{LS_ZPD} (right part of Fig.
\ref{fig:view}), the displacement field of the mode is a sum of the
rigid small-amplitude rotation and irrotational quadrupole term (the
latter provides vanishing velocity of electrons at the surface):
\begin{equation}
{\vec u}({\vec r})={\vec \Omega}\times {\vec r}+ \delta_2
(1+\delta_2/3)^{-1}%
{\bf \nabla}(yz)
\label{displ}
\end{equation}
where $\Omega$ is the angular velocity and $\delta_{2}=3/2 \;
(R^2_{\|}-R^2_{\bot})/(R^2_{\|}+2R^2_{\bot})$ is the parameter of
quadrupole deformation expressed via the semi-axes of the ellipsoidal
system.  Both macroscopic treatments, \cite{LS_ZPD,Iu_M1}, include the
rigid rotation of valence electrons versus the ions with the restoring
force originating from the Coulomb interaction between the electrons
and ions. But the model \cite{LS_ZPD} has the additional irrotational
quadrupole term. Its physical sense becomes clear if we take into
account that the SM, like the quadrupole E2 modes, is separated into
low-energy ($\Delta {\cal N} =0$) and high-energy ($\Delta {\cal N}
=2$) branches. The first rotational term in (\ref{displ}) is
responsible for the $\Delta {\cal N} =0$ branch while the second
irrotational term generates the $\Delta {\cal N} =2$ branch. Thus the
branches have quite different physical nature. This feature is
exploited as a signature of the Bose-Einstein condensate in dilute gas
of trapped Bose atoms \cite{BE_M1,BEexp_M1}. Unlike the normal phase
which exhibits both SM branches, the condensate phase supports only
the irrotational flow. So, a vanishing $\Delta {\cal N} =0$ branch
serves as a fingerprint of the condensate.

The rotational low-energy branch $\Delta {\cal N} =0$ corresponds to a
large extend to the scissors scenario exhibited in left part of Fig.
\ref{fig:view} and thus is usually treated as a true SM. In atomic
clusters, just this branch is responsible for van Vleck paramagnetism
and other effects of orbital magnetism. In what follows, we will
concentrate on this part of SM.

In axial clusters, the low-energy SM is generated by the orbital
momentum fields $L_x$ and $L_y$ perpendicular to the symmetry axis
$z$. The mode is represented by the states $|\Lambda^{\pi}=1^{+}>$
where $\Lambda$ is the eigenvalue of $L_z$ and $\pi$ is the space
parity. Energy and magnetic strength of the mode are estimated as
\cite{LS_M1,LS_ZPD,prl_M1,M1_ne_EPJD_03}
\begin{equation}
\omega = \frac{20.7}{r_{s}^{2}}N_{e}^{-1/3}\delta_{2} \ eV,
\label{eq:om}
\end{equation}
\begin{equation}
\label{eq:B(M1)}
B(M1)\; =  \; 4\langle 1^{+}\mid {\hat L}_{x}\mid 0\rangle^2\mu_{b}^{2}
\; = \; \frac{2}{3}N_e \overline{r^2} \omega \mu _{b}^{2}
\; \simeq \; N_{e}^{4/3}\delta_{2} \ \mu _{b}^{2}
\end{equation}
where $N_e$ is the number of valence electrons, $r_s$ the Wigner-Seitz
radius (in $\AA$), and $\mu_b$ is the Bohr magneton. We use here
atomic units $m_e=\hbar =c=1$. The value $B(M1)$ stands for summed
strength of the degenerated x- and y-branches. The z-branch vanishes
for symmetry reasons. It is worth noting that $B(M1$ transition
strength does not depend on $r_s$ and so is the same for different
metals. Both energy and magnetic strength are proportional to the
deformation parameter $\delta_{2}$ and thus vanish in spherical systems. So, the SM
exists only in system with a broken spherical symmetry.  The larger
the deformation, the stronger the low-energy scissors mode.

It worth noting that we assume here global deformation of the system.
At the same time, spherical symmetry can be broken locally while
keeping the global spherical shape. This takes place, e.g., in
spherical clusters where the ionic background as such destroys locally
the spherical symmetry and gives rise to some M1 (though rather weak)
orbital strength even with {\it zero} global deformation
\cite{Re_PRA_triax}.

The irrotational $\Delta {\cal N} =2$ branch of SM takes place in both
systems, spherical and deformed. Its energy maps the energy of the
quadrupole plasmon. In fact the irrotational SM is a part of this
plasmon.

\subsubsection{Microscopic view}

The microscopic treatment of the SM is based on the shell structure of
axially deformed mean field \cite{prl_M1,M1_ne_EPJD_03}. The angular
momenta orthogonal to the symmetry axis, ${\hat L}_{x}$ and ${\hat
  L}_{y}$, promote low-energy $\Delta {\cal N}=0$ transitions inside
the valence shell and high-energy $\Delta {\cal N}=2$ transitions
across two shells.

It is instructive to expand the wave functions
of the single-electron states in terms of the spherical basis $(n L\Lambda )$
\begin{equation}
 \Psi_{\nu =[{\cal N}n_z\Lambda ]}
 =
 \sum_{nL} a^{\nu}_{nL}R_{nL}(r) Y_{L\Lambda}(\Omega).
\end{equation}
This allows to evaluate the single-particle orbital M1 transition
amplitude between hole ($\nu =h$) and particle ($\nu =p$) states:
\begin{equation}\label{eq:me}
 \langle\Psi_{p}|{\hat L}_{x}|\Psi_{h}\rangle \propto
\delta^{\mbox{}}_{\pi_{p},\pi_{h}}\delta^{\mbox{}}_{\Lambda_{p},
\Lambda_{h}\!\pm\!1}
\sum_{nL}
a^{p}_{nL}a^{h}_{nL}\sqrt{L(L\!+\!1)\!-\!\Lambda_h(\Lambda_h\!\pm \!1)}.
 \nonumber
\end{equation}
Eq. (\ref{eq:me}) shows that the scissors mode is generated by
$\Lambda_p=\Lambda_h\pm 1$ transitions between the components of
the same spherical $(nL)$-level. In spherical systems
$(nL\Lambda)$-states belonging to the level $(nL)$ are degenerate
while in deformed systems they exhibit the deformation splitting and
so may be connected by M1 transitions with non-zero excitation
energies. This is the microscopic origin of the scissors mode.  The energy scale
of the scissors mode is determined by the deformation energy splitting
and so is rather small. This explains the predominantly low-energy
($\Delta \cal{N}$=0) character of the true SM.

There is an intimate connection between
the scissors and quadrupole E21 modes in deformed clusters.
If fact, both SM and E21  are parts of a general
motion characterized by the states  $|\Lambda^{\pi}=1^{+}>$.
To illustrate this point, let's again consider Eq. (\ref{eq:me}).
As was mentioned above, the SM operator connects only components
from one and the same basis level $nL$.
This is because the operators ${\hat L}_{x}$ and ${\hat L}_{y}$
have no $r$-dependent part and so, due to orthogonality
of the radial wave functions $R_{nL}(r)$, cannot connect the
components with different $nL$. But the latter can be done
by the quadrupole operator $r^2 Y_{21}$.
In this sense the SM operator is more selective than E21,
though both operators generate transitions of the same
multipolarity. The states $|\Lambda^{\pi}=1^+>$
involve both SM and E21 modes and respond to both M11 and E21
external fields. The states are treated as magnetic or electric, depending
on each of the two responses dominates.

\begin{figure}
%FIGURE 5
\centerline{\epsfig{figure=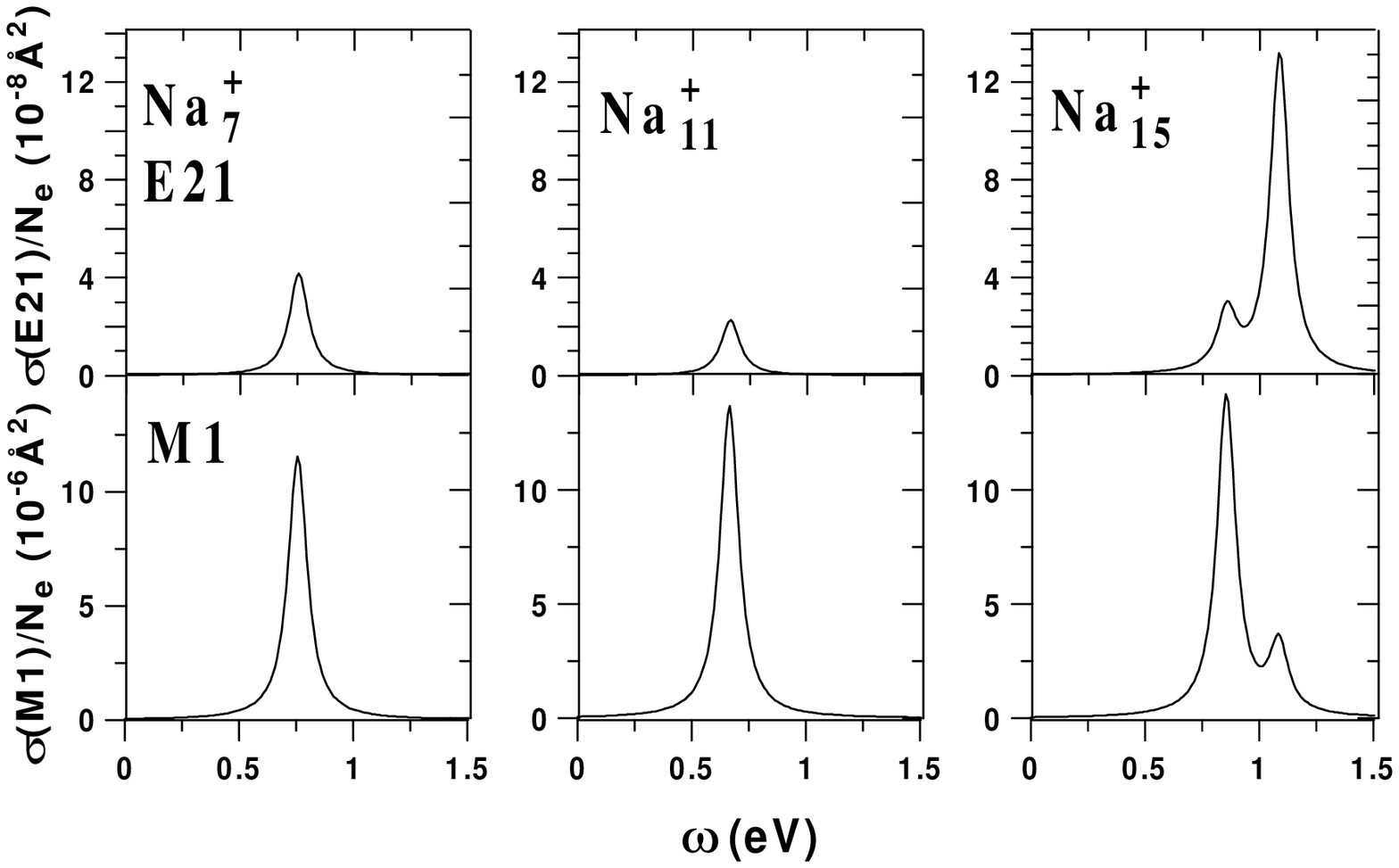,width=15cm,clip=}}
\caption{
\label{fig:E2_M1}\sl
E21 and scissors M1 photoabsorption for infrared modes in light
deformed clusters.}
\end{figure}

The dual nature of the states  $|\Lambda^{\pi}=1^{+}>$ is demonstrated in Fig.
\ref{fig:E2_M1}.  The one-to-one correspondence between the E21 and SM
peaks takes place.  These peaks represent $1eh$ states $\{
[101]-[110]\}_{21}$ in Na$^+_7$, $\{ [220]-[211]\}_{21}$ in
Na$^+_{11}$, and $\{ [211]-[200]\}_{21}$, $\{ [211]-[202]\}_{21}$ in
Na$^+_{15}$.  The first free states originate from the deformation
splitting of spherical levels and thus give rise to SM. In principle,
these states carry both E21 and SM flows and the ratio between two
contributions depends on the actual state structure.  The plot for
Na$_{15}^+$ gives an instructive example. It is seen that the lower
peak determined by $[211] \to [202]$ transition between the members of
the deformation multiplet exhibits an appreciable magnetic dipole
strength and thus should carry a large SM fraction. Just such
deformation-induced infrared states provide access to SM. Instead, the
higher peak is determined by a $[211] \to [200]$ transition which
takes place even in spherical case.  The deformation is not crucial
here. This state favors the E21 field and so can be treated as an
ordinary quadrupole mode.

\subsubsection{Effects of orbital magnetism}

Static orbital magnetism in clusters was widely explored during last decades.
The studies were mainly based on the Landau theory of the atomic magnetism
\cite{Landau}. A variety of issues was covered: giant dia- and
paramagnetism  in weak magnetic fields \cite{giant_dia,Kre},
size and temperature effects \cite{Ruit,Fra_su}, manifestation
of quantum supershells in magnetic susceptibility \cite{Fra_su},
influence of cluster shape (both axial and triaxial) on the
magnetic properties \cite{LS_ZPD,SRL}, anisotropy of magnetic
susceptibility in deformed clusters \cite{LS_ZPD,SRL},
orbital magnetism of supported clusters \cite{Binns}, etc..

In the present study, we will consider some principle points of the
orbital magnetism, connected with the scissors mode: the decisive role
of SM in van Vleck paramagnetism \cite{LS_ZPD,M1_ne_EPJD_03} and the
related effect of dia-para anisotropy in magnetic susceptibility of
particular light clusters \cite{M1_ne_EPJD_03}.  The latter effect
displays a peculiarity of small clusters to exhibit strong
variations in their properties with a cluster size.  For example, RPA
calculations \cite{prl_M1,M1_ne_EPJD_03} show that the SM energies and
B(M1) strengths, though mainly scaling with the deformation $\delta_2$
and the electron number $N_{\rm e}$ according to the trends
(\ref{eq:om}) and (\ref{eq:B(M1)}), demonstrate, nevertheless, strong
fluctuations. This can affect the magnetic susceptibility and lead, in
particular cases, to dia-para anisotropy.

At this point, we introduce a few equations to make the discussion more precise.
The interaction of cluster valence electrons with a uniform
magnetic field $B_k$ applied along the coordinate axis $k$
is
\begin{equation}
\label{eq:magn_int}
  {\hat H}_{int}= \mu_b B_k {\hat L}_k
  + \frac{1}{2}\mu_b^2 B_k^2 \rho^2_k
\end{equation}
where $k = x,y,z$ is the coordinate index,
${\hat L}_k=\sum_{a=1}^{N_{\rm e}} {\hat L^{(a)}}_k$ is
the $k$-th projection of the angular moment operator
(the sum runs over all valence electrons),
$\rho^2_z=\sum_{a=1}^{N_{\rm e}} (x_a^2+y_a^2)$,
$\rho^2_{x}=\sum_{a=1}^{N_{\rm e}} (y_a^2+z_a^2)$,
and $\rho^2_{y}=\sum_{a=1}^{N_{\rm e}} (x_a^2+z_a^2)$.
We neglected in (\ref{eq:magn_int}) electron spins
since for clusters considered below (axial sodium clusters
with even $N_{\rm e}$ and completely filled Fermi level)
their contribution to the magnetic susceptibility is expected to
be small.

If the magnetic field is weak, one can use the perturbation
theory. Then, up to the second order to $B_k$, the induced
change of the ground state energy is
\begin{eqnarray}
\label{eq:pert}
  \omega^{int}_0&=&
  \mu_b B_k <0|{\hat L}_k|0>
  -\mu_b^2 B_k^2\sum_{j\ne 0}
  \frac{|<j|{\hat L}_k|0>|^2}{\omega_{j}}
  \nonumber \\
  &+&\frac{1}{2}\mu_b^2 B_k^2 N_e \overline{\rho^2_k}
\end{eqnarray}
where $\omega_{j}$ is the energy of the excited state $|j>$
and $\overline{\rho^2_k}$ is the average value of $\rho^2_k$.
The negative second and positive third terms in
(\ref{eq:pert}) are responsible for the temperature
independent van Vleck paramagnetism and Langevin
diamagnetism, respectively.

The first (linear) term in (\ref{eq:pert}) dominates in the systems
with a partly filled Fermi level since in this case there is only
incomplete mutual compensation of the contributions of the valence
electrons with different orbital projections $\Lambda$. In particular,
this term results in a strong diamagnetic moments $\mu=\mu_b |\Lambda
|$ in axial clusters with odd $N_{\rm e}$ \cite{SRL}. In magic
spherical clusters, where the Fermi level is fully occupied, both the
first linear and second quadratic terms are zero and these clusters
are again diamagnetic \cite{SRL}. This diamagnetism is called giant
\cite{Kre} since, due to $\overline{\rho^2_k} \gg a^2_0$ ($a_0$ is the
Bohr radius), it is much stronger than the atomic one.

We will consider axially (z-symmetric) deformed clusters
with even $N_e$ and fully occupied Fermi
level. In this case, the linear term in
(\ref{eq:pert}) is zero but
there remains the van Vleck term for
$k=x,y$. The orbital magnetic susceptibility
$\chi_k =-\partial^2 \omega^{int}_0 /\partial B^2_k$
is then the sum of Langevin diamagnetic and van Vleck
paramagnetic terms:
\begin{equation}
\chi_k =\chi^{dia}_k + \chi^{para}_k,
\label{ms}
\end{equation}
where
\begin{equation}
\chi^{dia}_k =- \mu_b^2 N_e \overline{\rho^2_k>} = -\mu_b^2
\Theta_k^R,
\end{equation}
\begin{equation}
\label{eq:chi_para}
\chi^{para}_k=2\mu_b^2\sum_{j\ne 0}
\frac{|<j|{\hat L}_k|0>|^2}{\omega_j}
= \mu_b^2 \Theta_k,
\end{equation}
having denoted by
\begin{equation}
\label{eq:mom_in}
\Theta_{k}=2\sum_{j\ne 0}
\frac{|<j|{\hat L}_{k}|0>|^2}{\omega_j}
\end{equation}
the cranking moment of inertia and by
\begin{equation}
\Theta_{k}^R =N_e \overline{\rho^2_k}
\end{equation}
its rigid moment of inertia.

\begin{figure}
%FIGURE 6
\centerline{\epsfig{figure=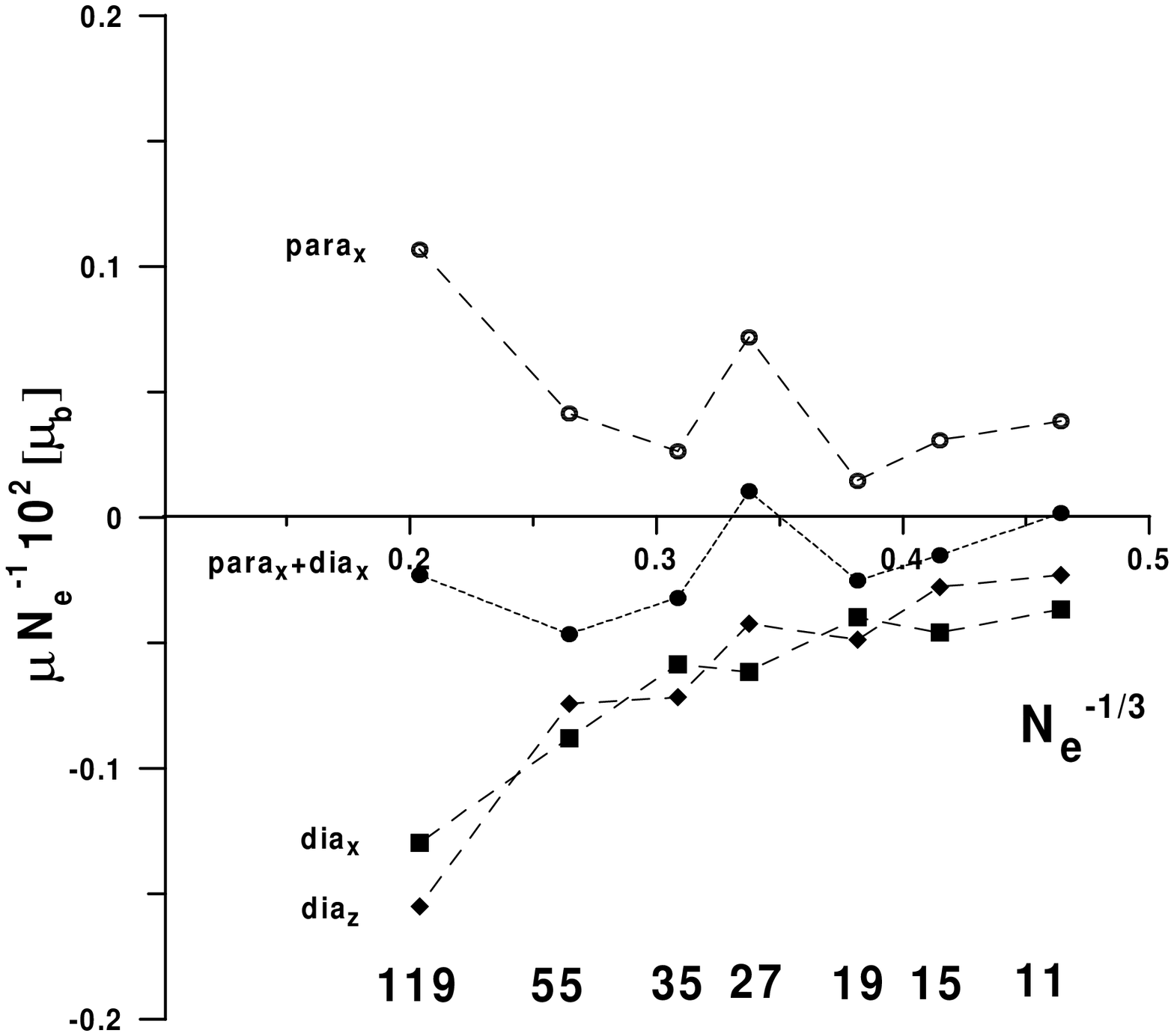,width=13cm,clip=}}
\caption{\label{fig:dia_para}\sl
Normalized diamagnetic, paramagnetic and summed moments
$\mu =\chi B_k$ ($B_k=4 T$) in axial deformed clusters
${\rm Na}_{11}^+$, ${\rm Na}_{15}^+$, ${\rm Na}_{19}^+$,
${\rm Na}_{27}^+$, ${\rm Na}_{35}^+$,
${\rm Na}_{55}^+$ and ${\rm Na}_{119}^+$.
}
\end{figure}

Note that for $k=x,y$ the operator entering in the matrix element in
(\ref{eq:chi_para}) is exactly the scissors generator. This makes evident
that just the {\it low-energy} SM mainly contributes to $\chi^{para}_{x,y}$.
Following our calculations \cite{M1_ne_EPJD_03}, this contribution achieves
$85 - 100\%$. So, just {\it the SM determines the van Vleck paramagnetism}.

In the schematic model
\cite{LS_ZPD}, the moment of inertia comes out as the rigid-body
value, so that $\theta_{x,y}=\theta_{x,y}^R$ and
$\chi^{para}_{x,y}= - \chi^{dia}_{x,y}$, i.e. a complete
compensation of dia- and paramagnetic terms in $\chi_{x,y}$ takes
place.  Due to axial symmetry, one also has $\chi^{para}_z=0$.  The
total susceptibility becomes, therefore, strictly anisotropic
\cite{LS_ZPD}
\begin{equation}\label{eq:bal}
\chi_x=\chi_y=0, \qquad \chi_z=\chi^{dia}_z,
\end{equation}
varying  from zero to diamagnetic values.

However, strong shell effects in some light clusters may alter
appreciably the above result.  This is illustrated for ${\rm
  Na}_{27}^+$ in Fig.  \ref{fig:dia_para}. Because of very low
excitation energy of the SM in this cluster
\cite{prl_M1,M1_ne_EPJD_03}, the paramagnetic susceptibility is so
much enhanced that it cannot be balanced by the diamagnetic term. So,
${\rm Na}_{27}^+$ gains the remarkable property to be paramagnetic in
x,y-directions and diamagnetic in z-direction.  The cluster ${\rm
  Na}_{11}^+$ also hints this property. Though dia- and paramagnetism
are weak as compared with other forms of magnetism, the dia-para
anisotropy described above is big enough to be measured
experimentally. This effect is only one of the examples of diverse
unusual magnetic properties of small clusters (see \cite{Binns} for a
recent review). Altogether, these properties may provide an
interesting perspective for new technologies.

\begin{figure}
%FIGURE 7
\centerline{\epsfig{figure=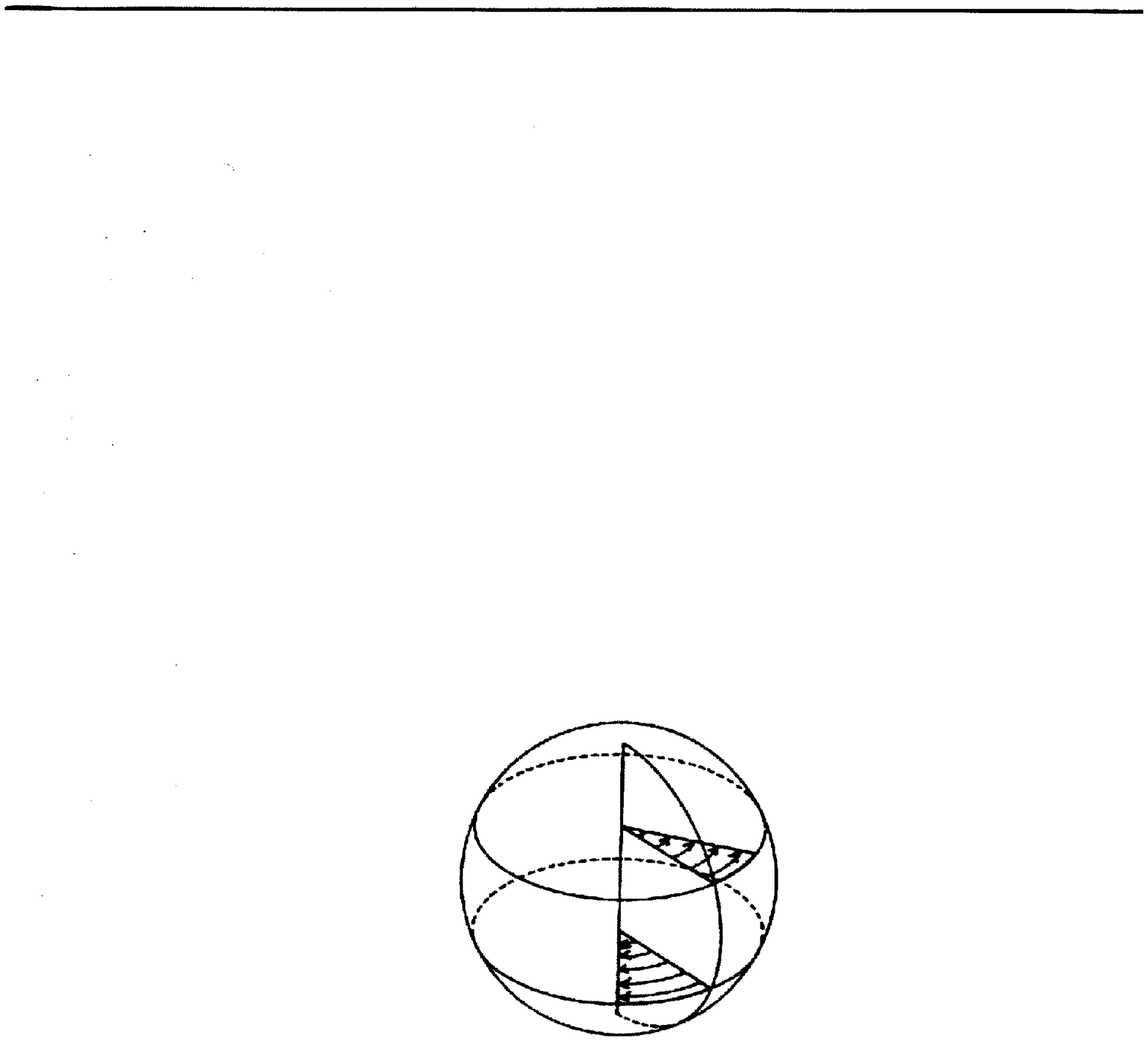,width=5cm,clip=}}
\caption{\label{fig:twist}\sl
Macroscopic image of twist mode.
}
\end{figure}

\subsection{Twist mode}

The SM is the dominant orbital magnetic mode in deformed clusters.
But in spherical clusters the SM vanishes and the next orbital magnetic mode,
twist M2, becomes the strongest one \cite{twist_prl_M2}.

This twist was first proposed as an quadrupole torsional vibrational
flow of an elastic globe \cite{Lamb}. The mode is generated by the
operator
${\hat T}=e^{-i\alpha z L_z}=e^{\alpha{\vec u}\cdot{\vec\nabla}}$
with the velocity field ${\vec u}=(yz,-xz,0)$ and small angle $\alpha$
\cite{Lamb,HE}.  The macroscopic image of twist is presented in Fig.
\ref{fig:twist}. The mode is treated as small-amplitude rotation-like
oscillations of upper hemisphere against the lower hemisphere. The
rotational angle of horizontal layers is proportional to $z$
(projection to the axis of rotation).  So, the flow vanishes at the
equator and poles of the system.

The restoring force of the twist is determined by the quadrupole
distortions of the Fermi surface in the {\it momentum} space. So, the
twist represents transverse magnetic quadrupole oscillations of an
{\it elastic} medium, generated by variation of the kinetic-energy
density. The twist is a general feature of any 3-dimensional finite
Fermi system which exhibits an elastic behavior. The twist mode is
well known in atomic nuclei and still remains an actual topic of both
theoretical and experimental studies \cite{HE}-\cite{Pon}. The
interest on twist is motivated by the fact that this mode is a
remarkable example of an elastic vortex motion. The twist was also
predicted but not yet observed in atomic clusters
\cite{twist_prl_M2,Bast} and trapped dilute gas of Fermi atoms
\cite{twist_FAT}.

Unlike the M1 scissor mode which exists only in deformed systems, the
twist takes place in Fermi systems of any shape, spherical and
deformed. As the magnetic quadrupole mode, the twist is represented by
$|\Lambda^{\pi}=2^->$ states of unnatural space parity. Twist energy and M2 strength
in atomic clusters are estimated as \cite{twist_prl_M2}
\begin{equation}  \label{eq:elast}
\omega = 17 eV \AA^2 r_s^{-2} N_e^{-1/3}, \quad B(M2)=0.52 r_s^2 N_e^2
\mu_b^2
\end{equation}
where B(M2) is the probability of M2 transition from the ground state
to $2^-$ twist state. It is easy to see that the generating operator
of the twist, $zL_z\propto r(Y_{10}L_z)$, coincides with the orbital
component $\mu=0$ of $M2$ transition operator \cite{BW}
\begin{equation}\label{M2_op}
{\hat{F}}(M2,\mu)=
\mu _b\sqrt{10}r[g_s\{Y_1{\hat{s}}\}_{2\mu}+\frac 23g_l\{Y_1{\hat{L}}%
\}_{2\mu}]
\end{equation}
which is the sum of spin and orbital contributions with $g_s=2$ and
$g_l=1$.  So, it is natural to consider the twist as a part of the
orbital M2 motion.

\begin{figure}
%FIGURE 8
\centerline{\epsfig{figure=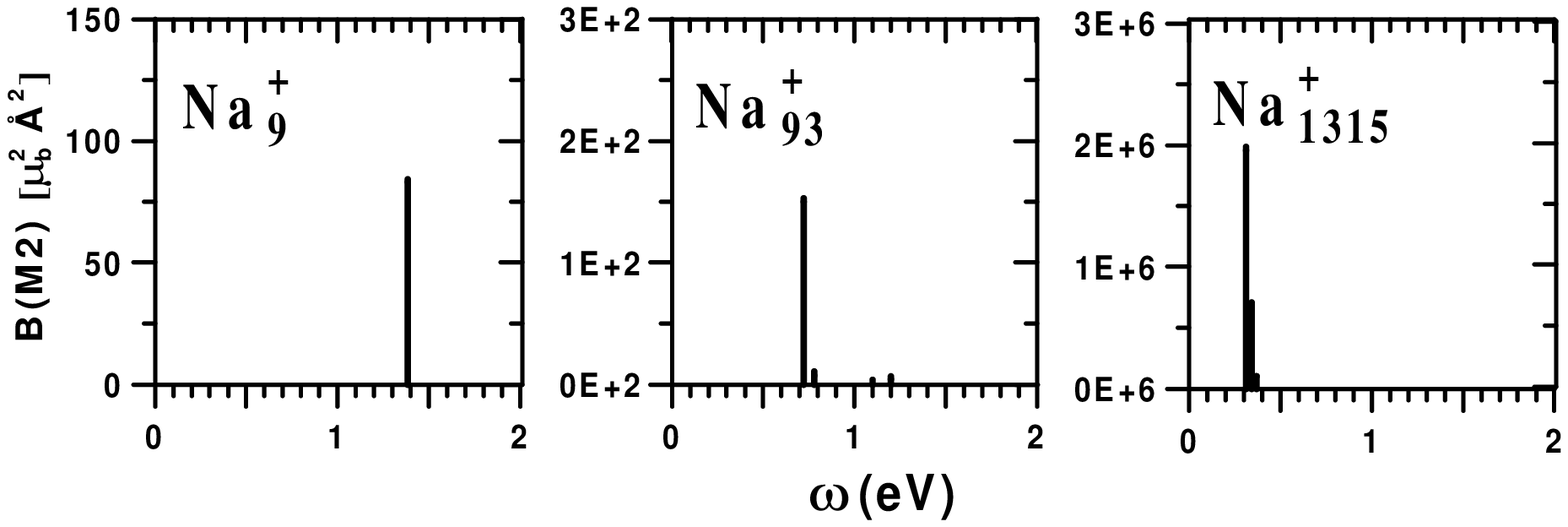,width=15cm,clip=}}
\caption{\label{fig:twist_eh}\sl
The distribution of $M2$-strength in spherical Na clusters of a different size.
}
\end{figure}
\begin{figure}
%FIGURE 9
\centerline{\epsfig{figure=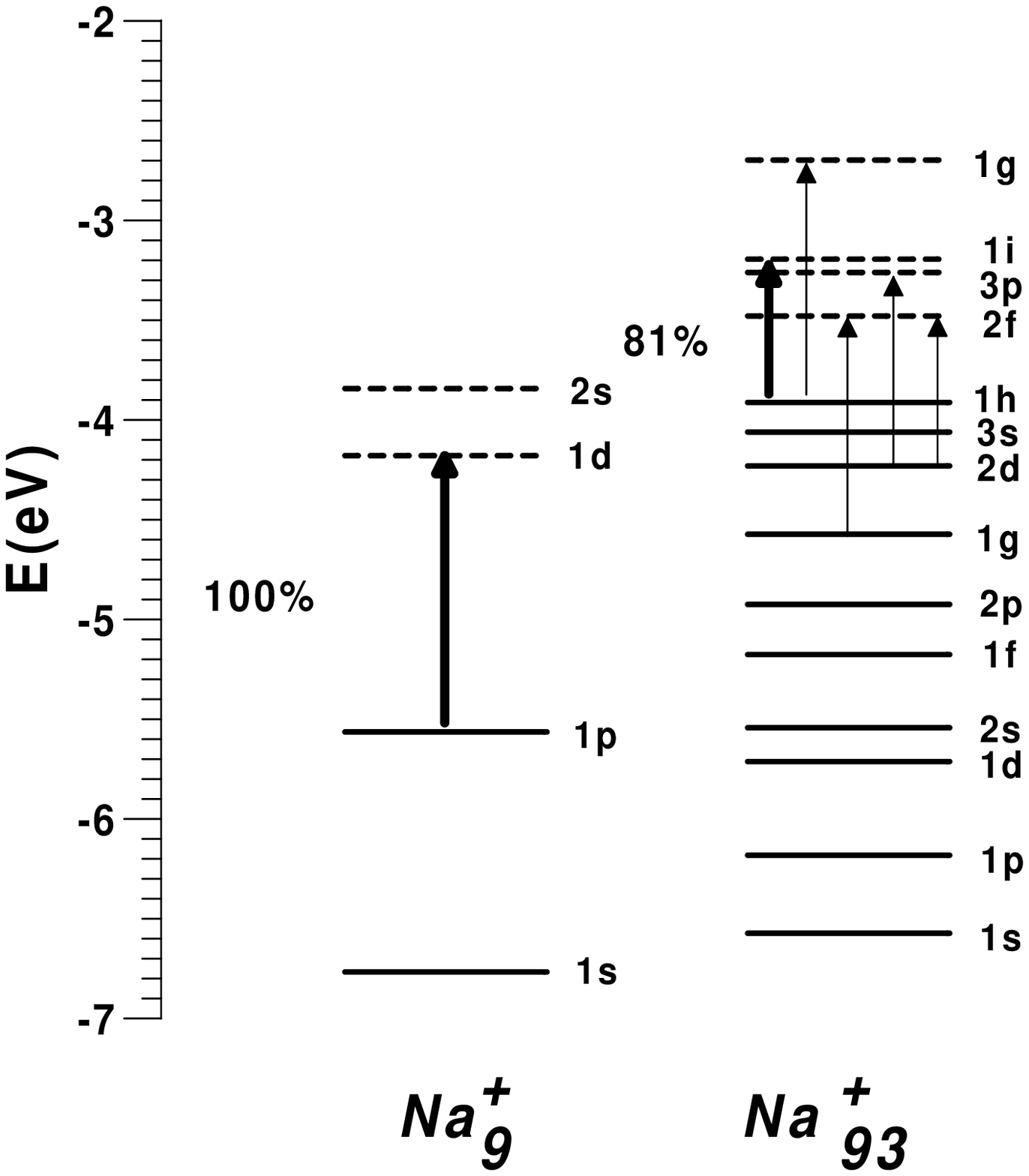,width=8cm,clip=}}
\caption{\label{fig:twist_levels}\sl
Single-particle level schemes for the mean field of valence electrons and M2 1h-1e
transitions in Na$^+_{9}$ and Na$^+_{93}$. The levels are labeled by $\{ nl \}$ where
$n$ is  the number of radial nodes in the single-electron function and $l$ is
the orbital moment. Transitions corresponding to the dominant twist peak are depicted by
bold arrows. Contributions  of these transitions to the complete $B(M2)$ strength
are given in $\%$.
}
\end{figure}

Let's consider now the microscopic features of the twist. Figure
\ref{fig:twist_eh} exhibits results of the microscopic calculations
\cite{twist_prl_M2} for the orbital M2 strength $B(M2)=|\langle
2^-|{\hat{F}}(M2)|0\rangle|^2$ in light, medium and heavy spherical
sodium clusters. Only the orbital part of the M2 transition operator
is used. It is seen that the dominant twist strength is concentrated
in a single 1eh peak with the lowest energy.  This peculiarity
persists independent of $N_e$, though the degree of concentration
decreases with $N_e$. The twist peak exhausts $100\%$, $80\%$, and
$60\%$ of the total M2 strength in $Na_9^{+}$, $Na_{93}^{+}$, and
$Na_{1315}^{+}$, respectively.  As was shown in \cite{twist_prl_M2},
this peak corresponds to the nodeless twist mode plotted in Fig.
\ref{fig:twist}.

The microscopic nature of the twist peak is exemplified in Fig.
\ref{fig:twist_levels}. It represents the $n,l \to n,l+1$ transition
between single-particle levels with the node number $n=1$ and {\it
  maximal} orbital moments $l$. The levels belong to the last occupied
and first empty electron quantum shells. As a result, the twist can in
principle deliver information about single-particle levels with {\it
  maximal} orbital momenta near the Fermi surface.

Calculations \cite{twist_prl_M2} show that the twist strength {\it
  dominates} over the spin strength already in clusters with $N_e=40$.
Starting with $N_e=92$, the orbital contribution becomes overwhelming
and demonstrates for $N_e=440$ a huge value of $2\cdot 10^5\mu
_b^2\AA^2eV$. As both spin and orbital long-wave M1 responses are
forbidden in spherical clusters (their M1-transition matrix elements
are proportional to the radial integral $\int
R_{n_1l_1}(r)R_{n_2l_2}r^2dr=\delta _{n_1l_1,n_2l_2}$ and thus vanish
in the non-diagonal case subject to ortho-normalization requirement),
the twist becomes the {\it strongest} multipole magnetic mode already
for clusters of medium size. This emphasizes its fundamental
character.

The search for the twist in atomic clusters is a demanding problem.
This mode cannot be observed in photoabsorption (only dipole) or
two-photon processes (only natural parity). Exploration of twist in
atomic nuclei hints that one might try to search this mode in clusters
by means of angular resolved electron-energy-loss-spectroscopy (EELS)
\cite{Ger,Fer,Eka}. Since twist contributes to the transverse
form-factor, the back angles of scattered electrons are most suitable.

\section{Experimental access to non-dipole modes}
%\label{sec:two_photon}
\label{sec:exp}

\subsection{General view}

As compared with other non-dipole modes, infrared quadrupole states
deserve special attention.  There are at least two samples, i) free
light deformed clusters and ii) oriented silver rods embedded in a
matrix, where these modes can be studied experimentally in two-photon
processes. The light deformed clusters are especially interesting
\cite{stirap_Ne_PRA} since their quadrupole modes provide access to
the single electron spectra. In turn, the spectra can serve as a
sensitive indicator of several cluster properties. It is known that
just small clusters demonstrate properties unusual for the bulk and,
in this sense, are attractive for both fundamental physics and
applications.

Two-photon processes (TPP) are widely used in atomic and molecular
physics (for a comprehensive discussion see \cite{Scoles}). Since
atomic clusters are similar to molecules, it would be natural to
implement the same reactions to clusters. However, applications of TPP
to clusters are still very scarce. This can be partly explained by our
poor knowledge on non-dipole low-energy electron modes in clusters,
and partly by peculiarities of clusters.

In this section we discuss applicability of some traditional TPP
methods to clusters with particular emphasis on infrared quadrupole
modes. In this connection, some cluster properties essential for TPP
should be closely scrutinized. Namely:

i) The dominant decay channel of cluster levels is usually not
radiative. The collective states, like the dipole plasmon, mainly
decay through Landau fragmentation (dissipation of the collective
motion through surrounding $1eh$ excitations).  For non-collective
states, the electron-electron collisions and electron-ion coupling are
most essential.

%%%%%%%%%%%%%%%%%%%%%%%%%%%%%%%%%%%%%%%%%%%%

ii) {\it Collective} electron modes in clusters have very short lifetimes. The
typical example is the dipole plasmon which represents in medium large clusters
with $N_e \sim 50-1000$ a broad ( $\sim$ 0.5-1 eV) mode with a life time of
10-20 fs.  This is due to a coupling of the collective strength with
energetically close 1eh states whose availability increases with increasing
cluster size \cite{rpaclust,rpasep}.  The mechanism is similar to Landau
damping of the plasmon in the bulk electron gas \cite{Lif88}. Its analogue in
finite systems is called therefore Landau fragmentation.  The short lifetimes
hamper application of adiabatic TPP methods as they would require more intense
laser irradiation.  The intense irradiation
%, on the other hand,
results in
strong dynamical Stark shifts and undesirable population of high-energy states
by second harmonic of the laser pulses.  These effects can be detrimental for
adiabatic TPP especially if the cluster spectrum is not sufficiently dilute.
{\it Non-collective} electron modes (as low-lying quadrupole $1eh$ ones)
have much longer life times that can approach ps for low temperature
and hundreds fs for the room temperature \cite{Cal,Reibook}.  Unlike
molecules, these modes in metal clusters are mainly driven by the
mean field similar to the nuclear one. This allows their unambiguous
treatment at least in light clusters with 5-20 atoms.

iii) Clusters exhibit a strong shape isomerism. Even beams with size
selected clusters usually represent a statistical mix of samples with
a multitude of different shapes. This blurs the measured low-energy
electron spectra.

The properties i)-iii) are absent in atoms but at least some of them are
common for
molecules.  Thus one may hope that methods of molecular spectroscopy
can be applied  to clusters. It is worth noting that, as compared
with molecules, clusters provide new possibilities for investigation
of electron modes. The electron spectra in clusters are very diverse
(from the collective plasmons to the pure $1eh$ configurations) and,
at the same time, are easily classified and treated in terms of
electronic quantum shells.

Light free clusters with 5-20 atoms have dilute spectra, a feature
which allows to avoid some of the troubles listed above. Landau
fragmentation in such clusters is very weak, the infrared levels are
not so broad and their lifetimes are much longer.  Most of light
clusters have larger isomer energies such that they have one clearly
preferred equilibrium shape and many of the energetically close
isomeric shapes are similar to the ground state one \cite{Kuemmel},
thus minimizing blurring of the low-energy spectra. Beams of
size-selected singly-charged light clusters are readily available. The
cluster temperature $\sim$ 100-200 K could be optimal for our aims. Then the
thermal broadening of electron levels is small.

Oriented silver rods embedded into glass \cite{rods_93,rods_98} represent
another interesting sample to be studied in TPP. These clusters are
known as commercial polarizers. The polarization is provided by the property
of the $\lambda\mu=10$ branch of the dipole
plasmon to absorb only the light aligned along the symmetry axis of
the cluster. Because of the big axis ratio, the rods represent an example of
atomic clusters with an extreme deformation. These heavy clusters can be
studied in Raman scattering  \cite{Duval}.

%%%%%%%%%%%%
% Figure 10
%%%%%%%%%%%%
\begin{figure} % Fig. 10
\centerline{\epsfig{figure=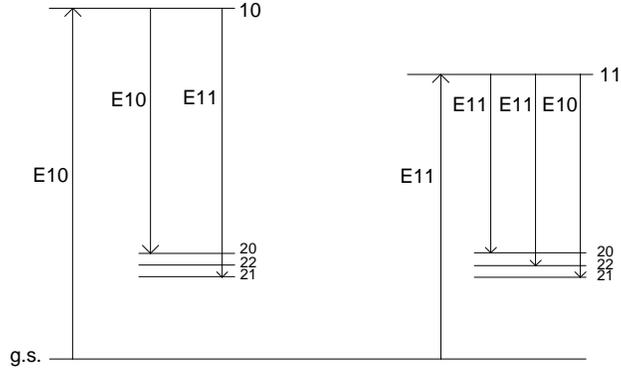,width=8.5cm,clip=}}
\caption{\label{fig:scheme}
Two-photon process: scheme of population of infrared quadrupole modes
$\lambda\mu =$20, 21 and
22 via the dipole states with $\lambda\mu =10$ (left) and 11 (right).
}
\end{figure}

Following the general TPP scheme presented in Fig. \ref{fig:scheme},
the target quadrupole level is populated via an intermediate dipole
state by two (absorption and emission) dipole transitions. Both the
collective dipole plasmon and non-collective $1eh$ configurations can
serve as intermediate states.  The TPP via the dipole plasmon, with
its broad and smoothed structure and ultra-fast decay through Landau
damping, reminds the population transfer via continuum.
%Because of a appreciable leaking from such intermediate media, TPP of this kind are
%generally not effective. However, they are possible
%in principle (see successful applications for the dipole plasmon in clusters
%\cite{Duval_PRB_01} and continuum in He atoms \cite{Yat}).
The TPP via non-collective $1eh$ dipole states is more transparent and
represents the typical population transfer process in
tree-bound-state $\Lambda$-systems.

As is shown in Fig. \ref{fig:scheme}, there are possible TPP
paths via dipole states with $\lambda\mu =10$ and 11. These states
could be, for instance, branches of the dipole plasmon, separated by
the deformation splitting. In the first case, the population of
$\lambda\mu =$22 mode is forbidden and only 20 and 21 modes can be
targeted. This provides some selection when carrying out TPP
experiments.

\begin{figure} % Fig. 11
%\centerline{\epsfig{figure=fig10.eps,width=9cm,clip=}}
\includegraphics[scale=0.4]{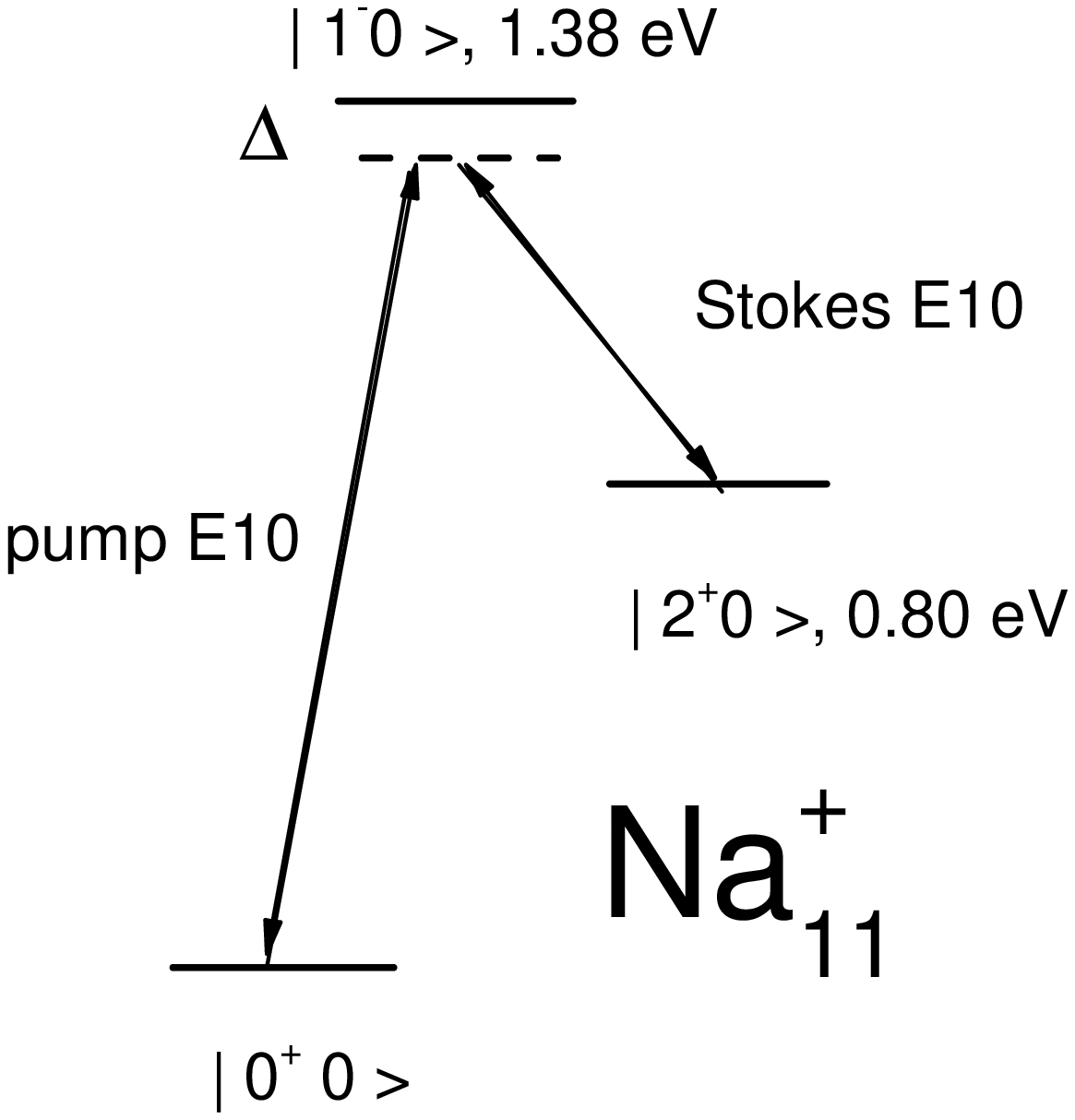}
\caption{\label{fig:sep_scheme}
TPP population transfer from the ground state $|0^+0>$ to the quadrupole state $|2^+0>$
via the $1eh$ dipole state $|1^-0>$ in Na$^+_{11}$. The detuning from the dipole state
is $\Delta \sim$ 0.14 eV.
%The similar TPP scheme is used for TPP
%via the dipole plasmon (see particular characteristics in Fig. \ref{fig:sep_plas}).
}
\end{figure}

For a crude analysis of TPP in clusters, one may use the
rotating wave approximation (RWA) that allows to discard the fast
laser frequencies and treat the problem with the much lower Rabi
frequencies only. Fig. \ref{fig:sep_scheme} for Na$^+_{11}$ shows a
typical electron $\Lambda$-system and hence the scale of the laser
frequencies corresponding to photoabsorption (pump) and photoemission
(Stokes) processes. Following our calculations, even for laser
pulses with intensity $I \sim 10^{10} W/cm^2$, Rabi frequencies in
light deformed sodium clusters lie in the interval $10^{-1} - 10^{-2}$
eV. Obviously, they are much less than the carrier (laser) frequencies
and thus fulfill the main RWA requirement.  Besides, because of the
pretty long life-time of the target non-collective quadrupole states,
one may operate with long laser pulses covering many Rabi cycles.
Hence we match the second RWA requirement.

%%%%%%%%%%%%
% Figure 11
%%%%%%%%%%%%

In the next subsections, we will consider some widespread two-photon processes
(Raman Scattering (RS), Stimulated Emission Pumping (SEP), and Stimulated Raman
Adiabatic Passage (STIRAP)) and estimate their ability to observe infrared
quadrupole modes.

\subsection{Raman scattering}

RS is one of the simplest two-photon processes where a dipole
laser-induced transition to an intermediate electronic level (real or
virtual) is followed by the dipole spontaneous fluorescence to low-energy levels.

RS measurements of electron infrared modes in clusters are yet very
rare. For instance, the quadrupole infrared modes were observed in
heavy silver clusters embedded into amorphous silica films
\cite{Duval_PRB_01}. The dipole plasmon was used as a resonant
intermediate state. In the recent experiment of this group
%The further measurements revealed that the modes are affected by cluster deformation \cite{Duval}.
with another sample (oriented silver rods embedded into glass), the
infrared electron quadrupole mode, probably $\lambda\mu =20$, was
detected \cite{Duval}. These observations show that the dipole plasmon
can be used as intermediate state for TPP in spite of its extremely
short lifetime. This is the encouraging message for TPP
implementation to clusters.

RS generally assumes that the coupling between the intermediate dipole
and target quadrupole states, from one side, and the coupling between
the intermediate and ground states, from the other side, are of the
same order of magnitude. Only then the final state is successfully
populated. This means that the absorption and emission dipole matrix
elements should be of the same scale. Experiments of E. Duval
\cite{Duval,Duval_PRB_01} show that this is indeed the case for heavy
clusters. In light deformed clusters the situation is more
complicated. Our calculations \cite{stirap_Ne_PRA} for Na$^+_7$,
Na$^+_{11}$ and Na$^+_{15}$ revealed that the absorption and emission
dipole matrix elements lie mainly in the interval 2-10 $ea_0$ (atomic
units) and thus are basically similar. This means that infrared
quadrupole modes in light deformed clusters have a chance to be
observed in the resonant RS running via the dipole plasmon.  However,
we have also found a few exceptions when the Stokes couplings are
strongly suppressed due to destructive interference effects (see
discussion in \cite{stirap_Ne_PRA}).  The particular example is oblate
Na$^+_7$ where only the path via the $\lambda\mu = 11$ branch of
dipole plasmon results in a strong TPP while the alternative path via
the $\lambda\mu = 10$ branch is about completely suppressed.  Such
considerable variations for different paths are not surprising for
small clusters where electronic spectra and transition rates are very
specific and can appreciably change from one sample to another. In
clusters with suppressed dipole couplings, the RS is not effective.
Besides, the RS has a general drawback that its photoemission step is
fully dictated by competition between different decay channels of
the intermediate state.  As a result, the spontaneous fluorescence to
the target state and hence the RS efficiency is typically rather low.
To overcome this trouble, we should consider methods with {\it
  stimulated} emission into the target level.

%
%%%%%%%%%%%
% Figure 13
%%%%%%%%%%%%
%\begin{figure}
%\includegraphics[height=13cm,width=9cm,angle=-90]{fig13.ps}
%\caption{\label{fig:sep_plas}
%%SEP population transfer from the ground state to the quadrupole state $|2^+0>$
%via the dipole plasmon in Na$^+_{11}$. See Fig. \ref{fig:sep_e1e2}
%for notations.
%}
%\end{figure}
%
%%%%%%%%%%%%%%%%%%%
\subsection{Stimulated emission pumping}

The stimulated emission pumping (SEP) enjoys widespread application
for atoms and molecules and seems to be very promising for population
of infrared quadrupole states in clusters. Unlike RS, this method
exploits two lasers, pump and Stokes (or dump) \cite{SEP,Vit_rew}. The
pump pulse is responsible for the first photoabsorption step. The
Stokes pulse couples the intermediate state to the target one. The
pulses can be simultaneous or sequential.  If the difference between
the pump and Stokes frequencies is in resonance with the frequency of
the target state, then the Stokes radiation stimulates emission to
this state.

One should distinguish SEP with incoherent and coherent irradiation.
If incoherent light sources are used, then, even in the best case of
the saturated process and coincident pump and Stokes pulses, the
maximal SEP efficiency achieves only one-third of the complete
population \cite{Vit_rew}. In practice, the transfer efficiency does
not exceed 10$\%$.  However, even this value is enough for many
spectroscopic studies.

A much better population can be achieved for coherent laser irradiation. If the
pump and Stokes pulses coincide, then under certain requirements (given in the
next subsection) we can even gain the fully adiabatic process (with still not
complete population transfer as compared with STIRAP \cite{Vit_98}). In this
subsection, we consider the coherent SEP with an appreciable detuning (often
called as off-resonant stimulated Raman).
%comparable with Rabi frequencies.
This process promises a simple, robust and efficient population scheme.

%%%%%%%%%%%%
% Figure 12
%%%%%%%%%%%%
\begin{figure}
\includegraphics[height=13cm,width=9cm,angle=-90]{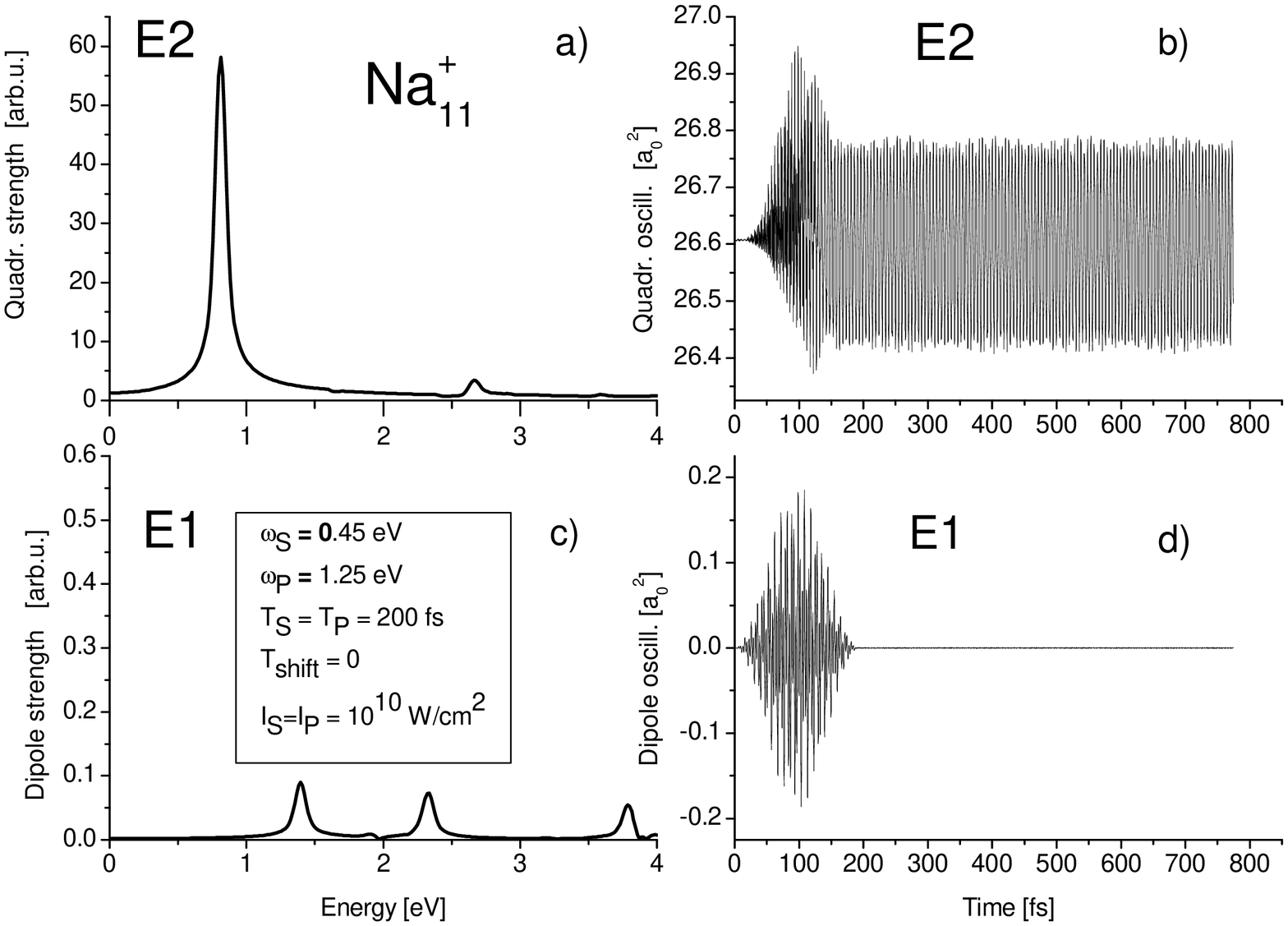}
%\centerline{\epsfig{figure=fig11.ps,width=9cm,clip=}}
\caption{\label{fig:sep_e1e2} Off-resonant SEP population transfer from the
ground state to the quadrupole state $|2^+0>$ via the $1eh$ dipole state
$|1^-0>$ in Na$^+_{11}$. The left plots exhibit (in arbitrary units) strengths
of quadrupole (a) and dipole (c) modes as a function of their energy. The right
plots depict (in atomic units) time evolutions of the overwhelming quadrupole
mode with $\omega_{2^+}$=0.80 eV (b) and dipole oscillations (d). Specifically,
the oscillations of the quadrupole and dipole moments are presented.
%It is seen that, unlike the dipole oscillations which survive only during
%the irradiation time, the quadrupole mode is very endurant.
The parameters of the Stokes (s) and and pump (p)
pulses (frequency $\omega$, duration $T$ and intensity $I$) are given
in the panel (c). The pulse profile is the squared cosine.
}
\end{figure}

We present below our TDLDA results for SEP in Na$^+_{11}$.
%(see Fig. \ref{fig:sep_scheme}).
%In Figs. \ref{fig:sep_e1e2} and \ref{fig:sep_e1e2_t}, we present our TDLDA results
%for this case.
The calculations explore the time evolution of the initial
single-electron states. All the possible channels including
photoionization (through absorbing boundary conditions) are taken into
account.  The ions are considered in soft jellium approximation at
fixed deformation. The details of the method are described elsewhere
\cite{Cal,Reibook}.  The parameters of the Stokes and pump pulses are
given the panel (c). Their frequencies maintain the two-photon
resonance with the target state: $\omega_p-\omega_s=\omega_{2^+0}$ and
are detuned somewhat below the intermediate dipole
state, $\Delta = (\omega_{1^-}-\omega_{0^+})-\omega_p
=(\omega_{1^-}-\omega_{2^+})-\omega_s =$ 0.14 eV, so as to minimize its
population.  It is worth noting that an appreciable detuning is
crucial as the dipole couplings in clusters are rather strong. This is
the only way to suppress undesirable population of other dipole and
quadrupole states and thus to enhance the population of the target
state.

In Fig. \ref{fig:sep_e1e2}, the TPP via the isolated non-collective
dipole level at 1.38 eV is considered (the actual $\Lambda$-system is
presented in Fig.  \ref{fig:sep_scheme}). The quadrupole ($\lambda=2$)
and dipole ($\lambda=1$) strengths %
$\sigma_{\rm E\lambda}(\omega)=\sum_j\langle j|r^{\lambda}Y_{\lambda\mu}|0\rangle^2
\delta(\omega_j-\omega)$
versus their excitation energy are depicted in the left panels.
Panel (a) demonstrates the
strong quadrupole excitation at 0.80 eV.  Because of the considerable
detuning, all other quadrupole and dipole modes are pretty small. The
right panels (b) and (d) show that, while dipole oscillations survive
only during the irradiation time $\sim$ 200 fs, the quadrupole
oscillations around the static values $Q_2=26.6 \; a_0^2$ ($a_0$ is
the Bohr atomic unit) are {\it very endurant}. There is no any
attenuation until 800 fs. Our study shows that the lifetime of this
mode is even longer and achieves at least a few ps. Because of the
dilute spectrum, decay of this mode via the electron-electron coupling
is negligible. Only the coupling with the ions remains but it is not
taken into account in these calculations. The electron-ion coupling
should limit lifetime of the mode by a few hundred fs at a room
temperature and by a few ps at a minimal temperature $\sim$10 K. The
latter case corresponds to our analysis.

%%%%%%%%%%%%%
% Figure 14
%%%%%%%%%%%%
\begin{figure}
\includegraphics[height=15cm,width=11cm,angle=-90]{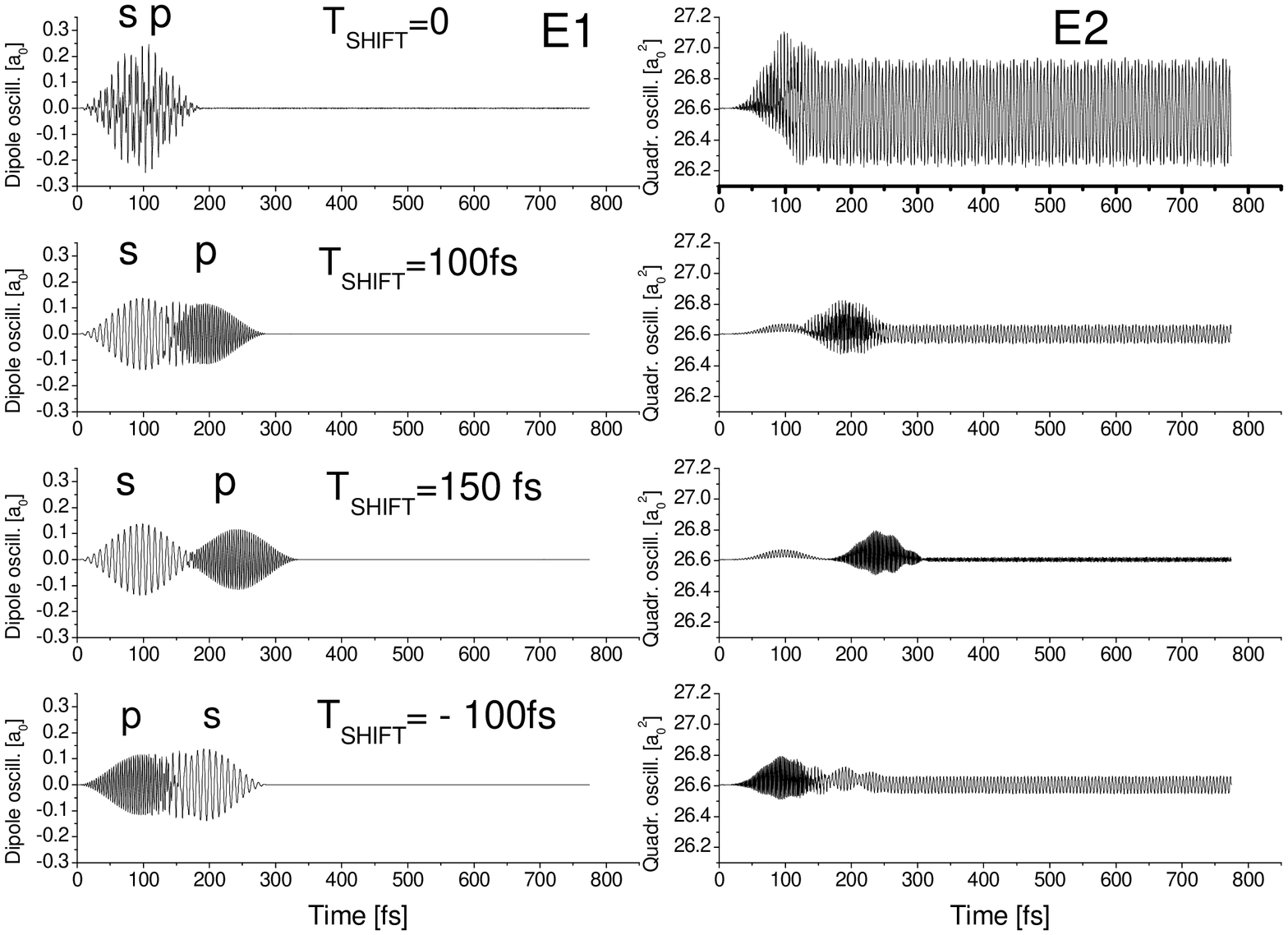}
%\centerline{\epsfig{figure=fig11.ps,width=9cm,clip=}}
\caption{\label{fig:sep_e1e2_t}
Time evolution of dipole and quadrupole oscillations for different pulse shifts (sequences)
in Na$^+_{11}$. The values of the shifts between the Stokes (s) and pump (p) pulses
are depicted at the left plots. The pulse parameters are the same as in Fig.
\protect\ref{fig:sep_e1e2} for exception of $I_s=3.3\cdot 10^{10}W/cm^2$.
}
\end{figure}

The calculations show that the endurant quadrupole mode is also
produced in TPP running through the dipole plasmon. In this case, the
mode is even several times stronger. The pulses with the frequencies
$\omega_P=$2.04 eV and $\omega_P=$1.24 eV and the same intensities and
durations as in Fig.  \ref{fig:sep_e1e2} were used. The pulses were
appreciably detuned from the dipole plasmon so as to weaken the competing
channels (compare $\omega_P=$2.04 eV with the dipole plasmon in
Na$^+_{11}$ depicted in Fig. \ref{fig:e1_plasmon}).  It is remarkable
that the dipole plasmon can be used as the effective mediator in spite
of its extremely short lifetime. The key point is the detuning that
prevents the population of the plasmon and thus the leaking from it.
Using the dipole plasmon as the intermediate state has the additional
advantage of dealing with the well know structure rather than with
still poorly known low-energy dipole states.

Fig. \ref{fig:sep_e1e2_t} demonstrates the dependence of the
quadrupole signal on the sequence of the pump and Stokes pulses. It
seen that the quadrupole mode is maximal at simultaneous pulses. This
is the typical SEP case. The process is rather robust to changes of pulse
characteristics. The calculations show that variations by several
times in pulse durations and intensities still maintain the stable and
distinctive SEP (though with the rescaled mode amplitude). Of course,
any change in the pulse intensities should be compensated by the
corresponding modification of the detuning.

It is a non-trivial problem to detect the quadrupole mode populated in
SEP.  In principle, this could be done by using the probe laser with
supplemented PES (photoelectron spectra) spectroscopy. The probe laser
should be intense enough to result in measurable PES and its frequency
should be large to map the single-electron states in the cluster and
thus provide the complete PES \cite{PESclust}. In Na$^+_{11}$, it could be
$\omega_{probe}\sim$ 7 - 8 eV. The probe pulse should follow the pump one
with a considerable delay so as to avoid overlapping with the latter
and map only the endurant quadrupole mode.
Due to the Raman coupling of the single-electron PES with the
quadrupole oscillation, every PES should acquire left and right
satellites with the relative frequencies coinciding with
$\omega_{2^+}$. Hence we get the frequency of the quadrupole state.
The strategy is to keep the fixed pump frequency with a reasonable
detuning bellow the intermediate dipole state and scan the Stokes
frequency. When we get the resonance satellites $\omega_P -
\omega_S=\omega_{2^+}$, then the endurant quadrupole mode has to be
produced and detected. It may happen that the Stokes frequency maps a
dipole state. In that case the endurant dipole oscillation are
accompanied by a subsequent considerable increase of photoemission
yield together with the appearance of dipole satellites. Such cases
can be easily separated from the case of interest, population of the
quadrupole mode, since the latter is not associated with a significant
resonant increase in photoemission yield but only by the satellites
signature alone.

The SEP experiments usually need three lasers: pump, Stokes and probe. However,
one may propose the simpler recipe with implementation of only pump and probe
lasers. Such scheme assumes the combination of the direct two-photon excitation
of the quadrupole state and PES techniques. Specifically, the intense pump
laser should scan until its two photons  excite the quadrupole state. Then the
delayed probe laser is applied in the manner described above to detect the
population. The calculations show that this simple method is quite effective
and robust.

%%%%%%%%%%%%%%%%%%%%
\subsection{STIRAP}

The STIRAP method \cite{Berg_Shore,Berg,Vit_rew} promises up to
100$\%$ population transfer from the initial to the target
level. This high efficiency is achieved by the {\it coherent adiabatic}
character of the process and the principle possibility to avoid any
leaking from the intermediate levels. Like SEP, STIRAP also implements
pump and Stokes lasers providing the couplings $|0> - |1>$ and $|1> -
|2>$, where $|0>$, $|1>$ and $|2>$ are initial ground state,
intermediate dipole state and target quadrupole states, respectively.
%The former couples the initial state $|0>$ with intermediate dipole state $|1>$
%and the latter couples the state $|1>$ with the target quadrupole state  $|2>$.
However, STIRAP is much more involved. Its main requirements are:

i) {\it Two-resonance condition} $\omega_P-\omega_S=\omega_2-\omega_0$
with a possible detuning
$\Delta=(\omega_1-\omega_0)-\omega_P=(\omega_1-\omega_2)-\omega_S$
from the intermediate state frequency.

ii) {\it Counterintuitive sequence} of partly overlapped  pulses when
the Stokes pulse comes before the pump one.

iii) {\it Adiabatic passage} of the state vector from the initial
to the target state. The adiabaticity condition (for the detuning $\Delta$=0)
has the form
\begin{equation}\label{adia}
  \Omega \Delta \tau >> 1
\end{equation}
where $\Omega=\sqrt{\Omega_p^2+\Omega_s^2}$ is the average of the pump
and Stokes Rabi frequencies and $\Delta \tau$ is the overlapping time of
the pulses (duration of the evolution process).

Under these requirements, one of the adiabatic time-dependent
eigen-functions of the system becomes a superposition of the initial
and target bare states only: $|b_0(t)>=c_0(t)|0>+c_2(t)|2>$. This is
so called {\it dark} state.  It is reduced to $|0>$ at the beginning
of the adiabatic evolution and to $|2>$ at the end.  Hence, the system
finally finds itself in the target state. The intermediate state $|1>$
is not involved in $|b_0(t)>$ and thus is not populated at any time.
Thus any leaking in the population is avoided and the transfer is
complete. The main point is to evolve the system adiabatically,
keeping it all the time in the state $|b_0(t)>$. STIRAP is well known
in atomic and molecular spectroscopy. It is rather insensitive to
precise pulse characteristics. Both continuous and pulse lasers can be
implemented.

Let's now examine the STIRAP requirements for the case of infrared
quadrupole modes in atomic clusters

{\it Two-resonance condition}.  The better fulfilled this condition,
the less mixing of the dark adiabatic state $|b_0(t)>$ with the state
$|1>$ and thus more robust the STIRAP. In practice, this condition is
perturbed by the detrimental dynamical (time dependent) Stark shifts
of the electron levels \cite{Vit_rew}
\begin{equation}
S_s(t)=-\frac{\Omega^2_s(t)}{4\Delta}, \qquad S_p(t)=-\frac{\Omega^2_p(t)}{4\Delta} .
\end{equation}
Too strong Stark shifts perturb the STIRAP balance and destroy the
adiabatic process.  The Rabi frequency reads \cite{Berg_Shore,Vit_rew}
\begin{equation}\label{Rabi_freq}
 \Omega = \frac{|d|}{\hbar}\sqrt{\frac{2I}{c\epsilon_0}}
 \simeq 2.20 \cdot 10^8 \; |d[ea_0]| \; \sqrt{I[\frac{W}{cm^2}]}
 \; s^{-1} \; ,
\end{equation}
where $d$ is the dipole coupling matrix element in atomic units, $I$
is the laser intensity in $W/cm^2$, $c$ is the light speed,
$\epsilon_0$ is the vacuum dielectric constant. It is seen that
$\Omega \sim \sqrt{I}$ and thus $S \sim I$. So, the smaller the laser
intensity, the weaker the Stark shift. However, the intensity cannot
be too small because of the adiabaticity condition (\ref{adia}). The
detuning $\Delta$ also has the upper limit established
%since it should not exceed
by the Rabi frequency $\Omega$.

Following our calculations, the coupling dipole matrix element in
light clusters is $d\sim 1 - 5 \; ea_0$. If to put the laser intensity
$I = 10^{10}$ $W/cm^2$, then Eq. (\ref{Rabi_freq}) estimates the peak
Rabi frequency as $\Omega \sim$ 0.02 - 0.1 fs or $\hbar\Omega  \sim$ 0.01 -
0.07 eV. These values are less than the detuning $\Delta$=0.136 eV
used for the SEP analysis in the previous subsection. Thus the SEP was
obtained at the tail of the dipole coupling between the states of the
$\Lambda$-system. Under the given detuning and laser intensity,
the Stokes shift is $S \sim$ 0.0004 - 0.01 eV which is still a bearable value.

{\it Counterintuitive order of pump and Stokes pulses}.  The order
when Stokes pulse precedes the pump one is crucial for the complete
population since only then the state vector can be fully projected
into the dark adiabatic state $|b_0(t)>$ \cite{Berg_Shore,Berg}.
Stokes and pump pulses must overlap and the overlapping time $\Delta
\tau$ determines the duration of the adiabatic evolution.

The $\Delta \tau$ should not be longer than the lifetimes of the
initial and target states while the lifetime of the intermediate state
does not matter if this state is not populated. So, the ultra-short
lifetime of the dipole plasmon (10-20 fs) is not the problem for the
ideal STIRAP. Even under some leaking from the plasmon (and thus the
modest coupling of the dark state $|b_0(t)>$ with $|1>$) STIRAP
is still possible. This is partly confirmed by recent experiment
\cite{Yat} where STIRAP was induced via continuum. Even in this
extreme case with ultra-fast decay channels and considerable leaking,
a clear STIRAP population transfer was obtained (though with low
efficiency $\sim 6\%$).

When STIRAP runs via an isolated non-collective dipole state, like
that with $\omega =$ 0.80 eV in Na$^+_{11}$, then the situation is
much simpler. Following the estimations for the lifetimes of $1eh$
state, the value of $\Delta \tau$ should not exceed a few hundreds fs
at room temperature and a few ps at low temperature.

It should be emphasized that the maximal population transfer at the
counterintuitive order of pump and Stokes pulses is the main signature
of STIRAP. Just this feature allows finally to judge if we deal with
STIRAP or other processes like SEP.

{\it Adiabatic passage}. This requirement is crucial for STIRAP. Only
adiabatic evolution can ensure the steady complete populations. The
adiabaticity condition is given by Eq. (\ref{adia}).  If we put,
following the above estimations, $\Omega \sim 0.02 - 0.1 fs$ and
$\Delta \tau \sim 300 fs$, then it is easy to see that the
adiabaticity condition is fulfilled: $\Omega \Delta \tau \sim 6 - 30
>> 1$.  However, the situation is not so simple since Eq. (\ref{adia})
does not take into account the detuning that effectively weakens the
Rabi frequency. In practice, we should use an appreciable detuning to
minimize the competing channels. But then $\Omega$ is considerably
decreased and the adiabatic condition is violated. This becomes a real
hindrance when the electron spectrum is not dilute enough.

It worth noting that the adiabaticity condition (\ref{adia}) can be
considerably relaxed.  Indeed, it was derived for {\it one}
intermediate level. But the realistic spectrum of the dipole plasmon
consists of a broad structure consisting from a sequence of dipole
levels (see Fig. 1). In this case, the STIRAP adiabatic condition
should be revised. The realistic case is more complicated but, at the
same time, opens new possibilities for STIRAP. In particular, one may
be allowed to loose the adiabatic STIRAP condition and thus the
requirements for the laser intensity.

A general case of N intermediate states, each with its own coupling
and detuning was studied in \cite{Vitanov}. It was shown that the
trapped adiabatic state $|b_0(t)>$ can be created only when the ratio
between each pump coupling and the respective Stokes coupling is the
same for all intermediate states. Following our calculations, this
condition is unrealistic for atomic clusters. However, softer
alternative adiabatic requirements can be formulated. In particular,
in the general case of arbitrary couplings, one may tune the pump and
Stokes lasers just below all intermediate states and thus form so
called adiabatic-transfer state which also results in a high, although
not complete, population of the target level. Unlike $|b_0(t)>$, the
adiabatic-transfer state can have admixture from the intermediate
states during the evolution period $\Delta \tau$ and so some
population leaking is unavoidable. Nevertheless, we have here a solid
adiabatic transfer with a high population of the target state.
Just this method is implemented in our calculations while using the dipole plasmon
as the intermediate TPP state.

%%%%%%%%%%%%
% Figure 15
%%%%%%%%%%%%
\begin{figure}
\includegraphics[height=14cm,width=10cm,angle=-90]{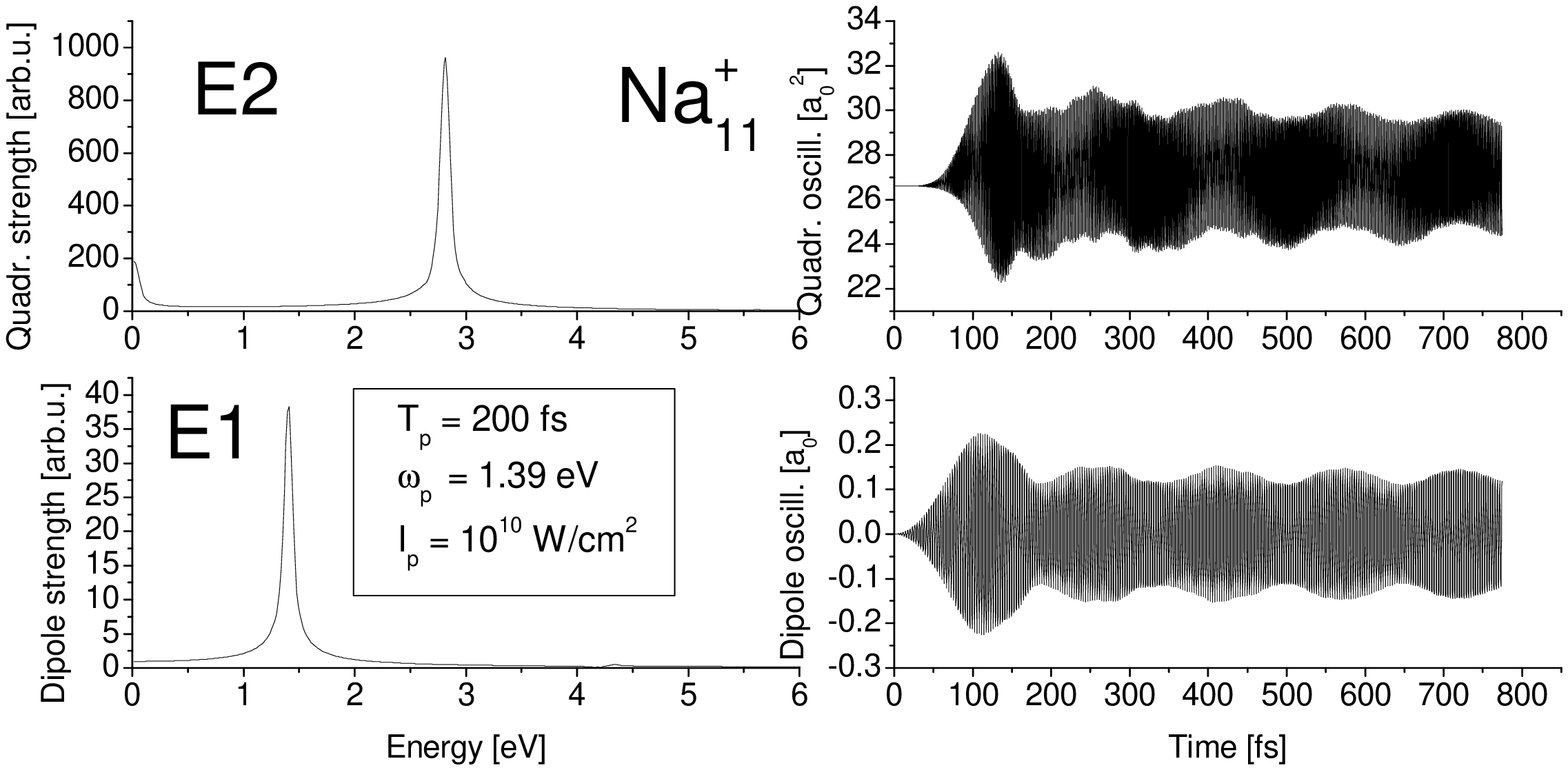}
%\centerline{\epsfig{figure=fig11.ps,width=9cm,clip=}}
\caption{\label{fig:e2_plas}
Simultaneous resonant excitation of the dipole state and quadrupole plasmon in
Na$^+_{11}$ by the first and second pump harmonics, respectively.
The up and middle left plots exhibit strengths of the quadrupole and dipole modes
as a function of their energy. The right plots depict time evolutions of these modes.
%The left-bottom plot shows the yield of photo-electrons (PES) as a function of their
%kinetic energy. The resonant ($\omega_P-\omega_{1-}=0$) and non-resonant
%($\omega_P-\omega_{1-}=0.136 eV$) cases are presented. In the first case, the PES
%signal is larger by several orders of magnitude and exhibits distinct peaks corresponding
%to $1s$, $1p$ and $1d$ mean field single-electron levels. The right-bottom panel
%presents parameters of the pump pulse.
}
\end{figure}

Altogether, the arguments and estimations presented above show that STIRAP is
possible in clusters. However, it is not so trivial to find the optimal
parameters of the process and get STIRAP in realistic TDLDA calculations.
Serious problems arise in connection with dipole and quadrupole plasmons which
reside in the energy region 2 - 4 eV and thus are always covered by multiphoton
excitations of the pump and Stokes pulses.  Hence there arises a considerable
undesirable population leaking. Especially strong coupling with the quadrupole
plasmon takes place when the pump or Stokes frequencies are in resonance with
some dipole state. The example of such process is exhibited in Fig.
\ref{fig:e2_plas} where the resonant excitation of the dipole state at 1.38 eV
leads to a huge population of the quadrupole plasmon. This means that we should
avoid the resonance cases by imposing appreciable detuning. But, as was
mentioned above, this will effectively weaken the Rabi frequencies.  Then, to
keep the requirement (\ref{adia}), one should use longer pulses. This is quite
possible at low temperatures when the lifetimes of the states can reach a few
ps.  In this case, we even have the freedom to decrease the laser intensity
with the aim to minimize the Stark shifts.

%%%%%%%%%%%%
% Figure 15
%%%%%%%%%%%%
\begin{figure}
\includegraphics[height=13cm,width=9cm,angle=-90]{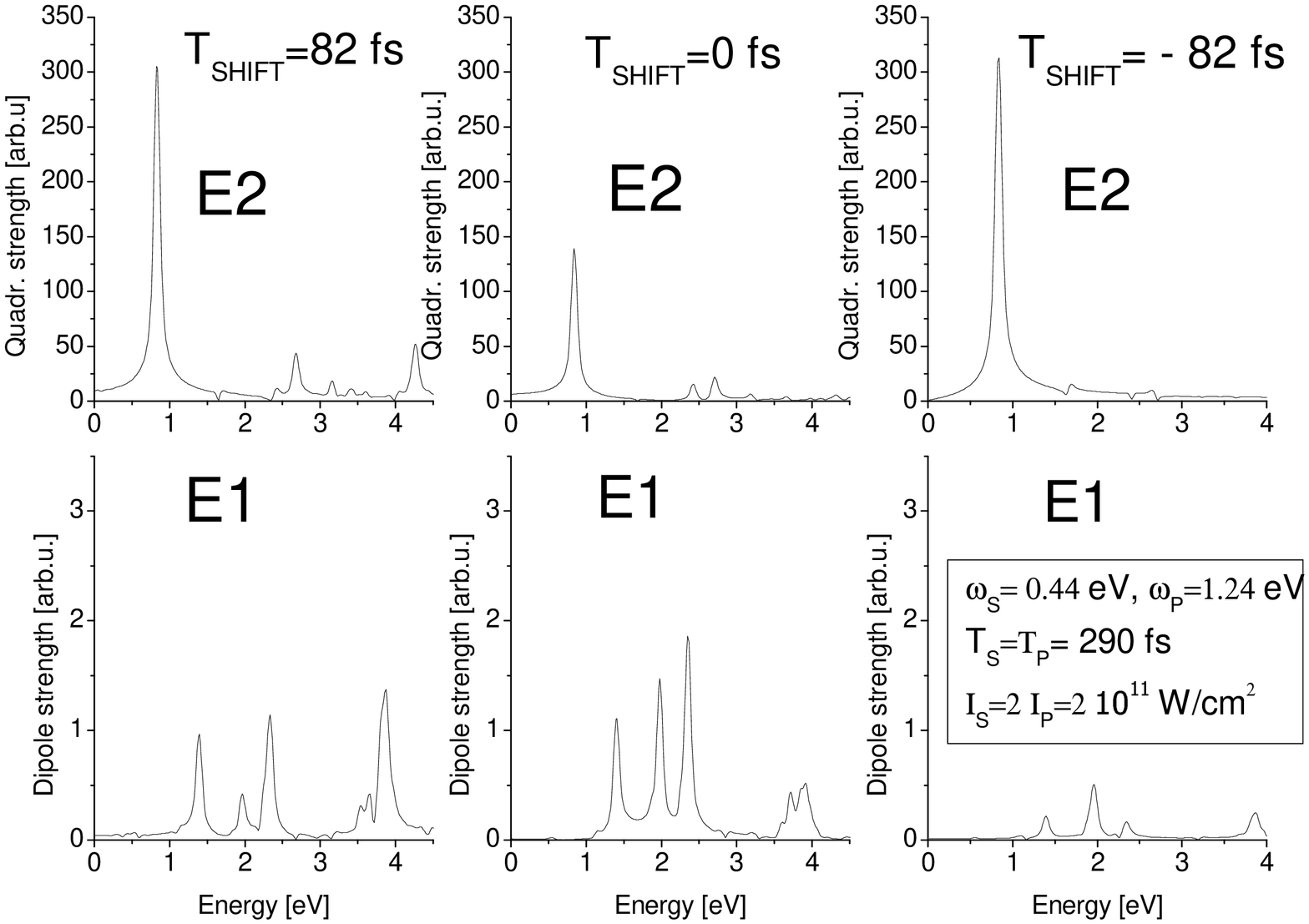}
\caption{\label{fig:strap}
Population transfer from the ground state to the quadrupole state $|2^+0>$
via the $1eh$ dipole state $|1^-0>$ in Na$^+_{11}$. The up and low panels
represent the quadrupole and dipole strengths at different delays (shifts)
of the pump pulse versus the Stokes one: 82 fs (counterintuitive STIRAP pulse sequence),
0 fs (coinciding pulses), -82 fs (intuitive pulse sequence).
The parameters of the Stokes (s) and and pump (p)
pulses (frequency $\omega$, duration $T$ and intensity $I$) are given
in the low-right panel.
}
\end{figure}

The proper choice of the pulse parameters allows to produce a STIRAP-like
process in clusters even at high laser intensity. A successful example is
presented in Fig. \ref{fig:strap}.  A large detuning of $\Delta =$ 0.15 eV is
here combined with a high pulse intensity of $I \sim 10^{11} W/cm^2$. This
particular choice allows to minimize the impact of the quadrupole and dipole
plasmons and, at the same time, to keep the adiabaticity condition. Fig.
\ref{fig:strap} shows that the counterintuitive pulse sequence provides the
stronger strength
%(and thus the population)
of the quadrupole mode than the coinciding pulses. In terms of the population,
this could be  the unequivocal signature of the STIRAP. As is seen from the
figure, the intuitive pulse sequence also results in the same quadrupole
strength.  It looks like this TPP embraces not only the typical STIRAP branch
corresponding to the dark state $|b_0>$ but also the branch from the dark state
$|b_{+}>$. The latter takes place because of the large detuning below the
intermediate dipole state. Folowing \cite{Gaubatz}, an appreciable detuning
below or above the intermediate state results in the additional dark dressed
state $|b_{+}>$ or $|b_{-}>$, respectively.

Though the process in Fig. \ref{fig:strap} looks like STIRAP, its STIRAP nature
still should be proved.
The point is that the strength exhibited in Fig. \ref{fig:strap}  does not correspond
to the population (squared amplitude of the target state) but rather to the
coherence (product of amplitudes of initial and target states). Hence,
the maxima of the strength at non-zero pulse shifts still do not mean
that the similar maxima take place for the population of the target state.
Our TDLDA model does not allow
to estimate properly the population. However, this can be done within the
Rotating Wave Approximation for 3-level $\Lambda$-system \cite{Vit_rew}.
Following these estimations, we have found that
the TPP in the example above  has the maximal population at
coinciding pulses and thus is not STIRAP. For getting the STIRAP we need more
intense laser pulses and maybe another value of the detuning.
One may get other, and more favorable, options in
clusters by a more suitable choice of the process parameters, set to worked
out. In any case, the example in Fig. \ref{fig:strap} hints the principle
possibility of STIRAP in atomic clusters.  The experimental detection of the
STIRAP population can be done by the same methods as in SEP.

The application of STIRAP to clusters seems to be generally feasible.
At the same time, this process is more demanding to realize than SEP. It
requires further
studies within realistic TDLDA models to look for the most effective process
configurations. STIRAP can hardly be more practicable than SEP
in population and detection of infrared quadrupole modes in clusters.
However, one may try to implement STIRAP to get more transfer efficiency
in particular cases. Besides,
STIRAP running via the dipole plasmon can be interesting for
theory of adiabatic TPP as such. Indeed, the intermediate state like
the dipole plasmon does not exist in atoms and molecules and so was
not yet investigated. Perhaps, this intermediate state will offer new
options for STIRAP and extend our knowledge about this intriguing TPP.

\section{Conclusion}
\label{sec:summary}

Non-dipole infrared electron modes in atomic clusters (electric
quadrupole (E2), magnetic dipole scissors (M1) and magnetic quadrupole
twist (M2)) were analyzed theoretically using as tool the
time-dependent local-density approximation (TDLDA). The collective and
non-collective nature of these modes as well as their connection with
the basic cluster properties (single electron spectra, orbital
magnetism) were analyzed in detail. Free light deformed sodium
clusters and embedded oriented silver rods were considered as the most
promising samples. The possible routes of experimental observation of
infrared quadrupole modes in two-photon processes (Raman scattering,
stimulated emission pumping (SEP) and stimulated Raman adiabatic
passage (STIRAP)) were scrutinized. Following our analysis, the
combination of SEP and PES techniques provides probably the most
simple and robust way to populate and detect the infrared quadrupole
states in free light deformed clusters. As a complementary method, a
technique using direct two-photon excitation was also considered.
The Raman scattering seems to be optimal for large clusters,
especially embedded oriented silver rods. The calculations encourage
applicability of STIRAP to clusters. However, this intriguing but demanding method
still needs further realistic TDLDA studies. Smaller clusters as, e.g.,
Na$^+_7$ are the most promising candidates.

We hope that our study will stimulate investigation of non-dipole
electron modes in clusters by means of TPP experimental methods. This
would provide a deeper understanding both clusters properties and
peculiarities of different TPP schemes.

\begin{center}
{\bf Acknowledgement}
\end{center}

V.O.N. thanks Professors K. Bergmann, E. Duval and J.-M. Rost and Junior
Professor T. Halfmann for useful discussions. The work was partly supported by
the Bundesministerium Bildung und Forschung (project No. 06ER124), the
Visitors Program of Max Planck Institute for the Physics of Complex Systems
(Dresden, Germany) and Heisenberg-Landau program (Germany - BLTP JINR).

\end{document}